\documentclass[lettersize,journal]{IEEEtran}
\usepackage{amsmath,amsfonts}
\usepackage{algorithmic}
\usepackage{array}
\usepackage[caption=false,font=normalsize,labelfont=sf,textfont=sf]{subfig}
\usepackage{textcomp}
\usepackage{stfloats}
\usepackage{url}
\usepackage{verbatim}
\usepackage{cite}
\usepackage[pagebackref,breaklinks,colorlinks]{hyperref}
\usepackage{graphicx}
\usepackage{bbding}
\usepackage{booktabs}
\usepackage{multirow}
\usepackage{float}
\def\BibTeX{{\rm B\kern-.05em{\sc i\kern-.025em b}\kern-.08em
    T\kern-.1667em\lower.7ex\hbox{E}\kern-.125emX}}
\usepackage{balance}

\usepackage{color,xcolor}
\usepackage{caption}
\usepackage{algorithm}
\usepackage[capitalize]{cleveref}

\usepackage[export]{adjustbox}
\crefname{section}{Sec.}{Secs.}
\Crefname{section}{Section}{Sections}
\Crefname{table}{Table}{Tables}
\crefname{table}{TABLE}{TABLES}

\newcommand\blfootnote[1]{%
\begingroup
\renewcommand\thefootnote{}\footnote{#1}%
\addtocounter{footnote}{-1}%
\endgroup
}

\begin{document}
\title{Toward Real World Stereo Image Super-Resolution via Hybrid Degradation Model and Discriminator for Implied Stereo Image Information}
\author{Yuanbo~Zhou\IEEEauthorrefmark{1},
	Yuyang~Xue \IEEEauthorrefmark{1},
    Jiang~Bi,
    Wenlin~He,
    Xinlin Zhang,
	Jiajun Zhang,
	Wei~Deng, \\
    Ruofeng~Nie,
    Junlin~Lan,
	Qinquan~Gao \Envelope,
	and Tong~Tong \Envelope
}

\markboth{Journal of \LaTeX\ Class Files,~Vol.~18, No.~9, September~2020}%
{How to Use the IEEEtran \LaTeX \ Templates}

\twocolumn[{%
	\renewcommand\twocolumn[1][]{#1}%
	\maketitle
	\begin{center}
		\def\s{0.1\linewidth}%
		\centering%
	
		\captionsetup{type=figure}
        \setlength{\tabcolsep}{2mm}
        \scriptsize
        {\begin{tabular}{cc}
                \hspace{-0.4cm}
                \begin{adjustbox}{valign=t}
                \begin{tabular}{c}
                \includegraphics[width=0.12\textwidth,height=0.125\textheight]{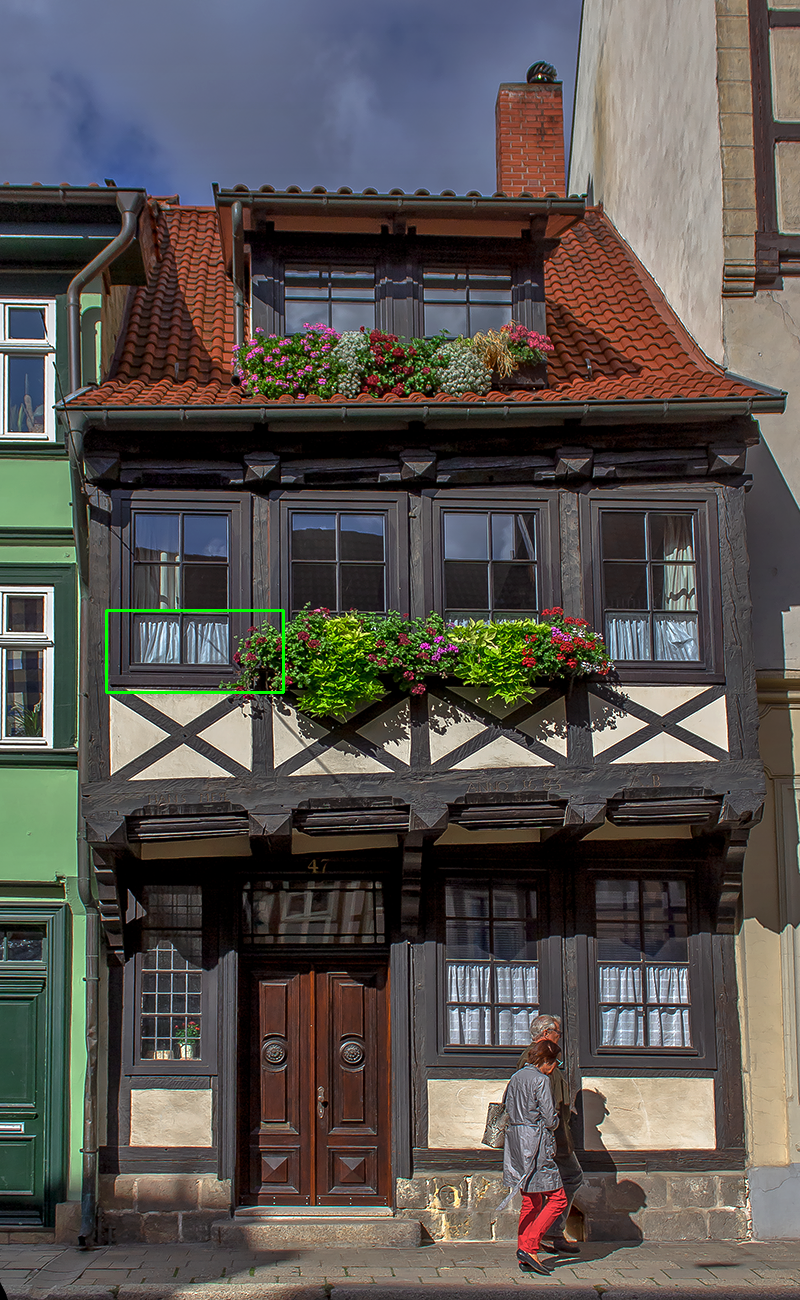}
                \\
                Left image
                \\
                \includegraphics[width=0.12\textwidth,height=0.16\textheight]{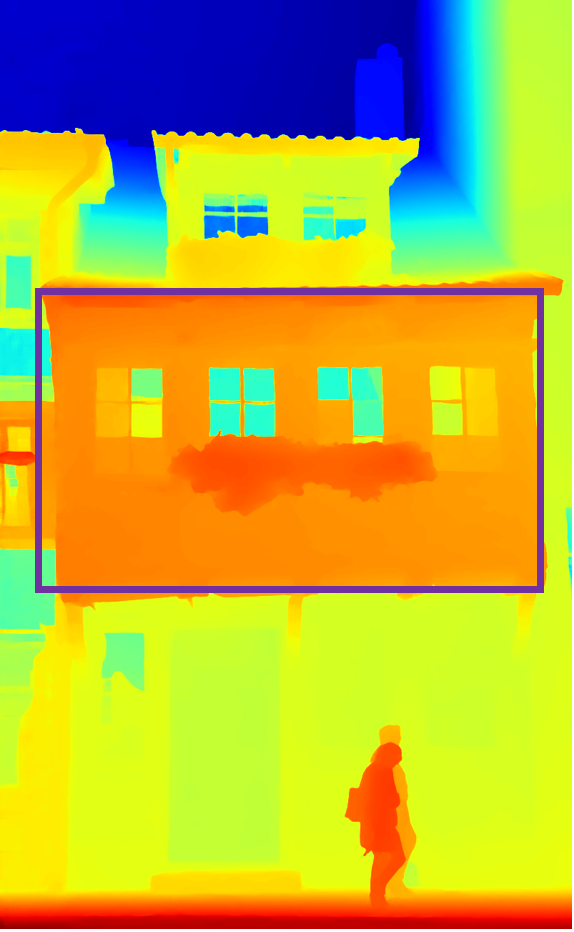}
                \\
                Disparity image
                \\
                \end{tabular}
                \end{adjustbox}
                \hspace{-0.46cm}
                \begin{adjustbox}{valign=t}
                \begin{tabular}{cccccc}
                \includegraphics[width=0.161\textwidth]{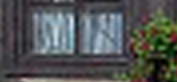} \hspace{-4mm} &
                \includegraphics[width=0.161\textwidth]{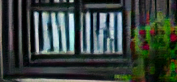} \hspace{-4mm} &
                \includegraphics[width=0.161\textwidth]{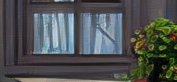} \hspace{-4mm} &
                \includegraphics[width=0.161\textwidth]{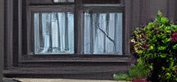} \hspace{-4mm} &
                \includegraphics[width=0.161\textwidth]{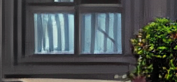} \hspace{-4mm}
                \\

                Bicubic \hspace{-4mm} &
                RealSR(CVPRW20')~\cite{ji2020real} \hspace{-4mm} &
                BSRGAN(ICCV21')~\cite{zhang2021designing} \hspace{-4mm} &
                SwinIR(ICCVW21')~\cite{liang2021swinir} \hspace{-4mm} &
                RealESRGAN(CVPR21')~\cite{wang2021real} \hspace{-4mm} &
                \\

                \includegraphics[width=0.161\textwidth]{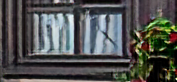} \hspace{-4mm} &
                \includegraphics[width=0.161\textwidth]{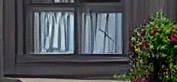} \hspace{-4mm} &
                \includegraphics[width=0.161\textwidth]{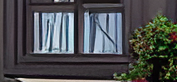} \hspace{-4mm} &
                \includegraphics[width=0.161\textwidth]{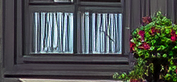} \hspace{-4mm}   &
                \includegraphics[width=0.161\textwidth]{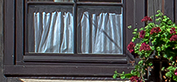} \hspace{-4mm}
                \\

                PDM-SRGAN(CVPR22')~\cite{luo2022learning} \hspace{-4mm} &
                DASR(ECCV22')~\cite{liang2022efficient}  \hspace{-4mm} &
                MMRealSR(ECCV22')~\cite{mou2022metric}  \hspace{-4mm} &
                RealSCGLAGAN(\textbf{Ours})    \hspace{-4mm} &
                GT \hspace{-4mm}
                \\

                \includegraphics[width=0.161\textwidth]{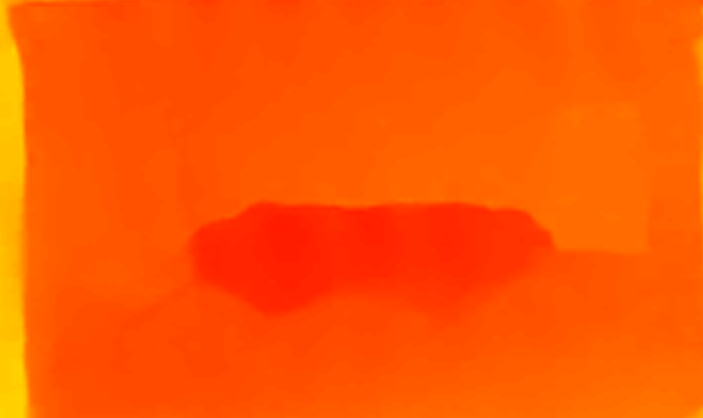} \hspace{-4mm} &
                \includegraphics[width=0.161\textwidth]{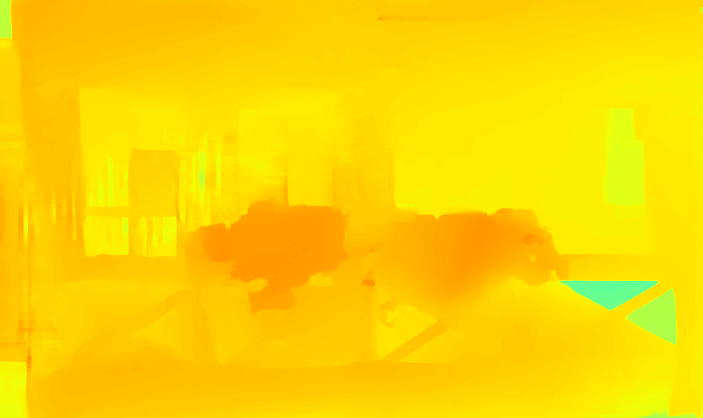} \hspace{-4mm} &
                \includegraphics[width=0.161\textwidth]{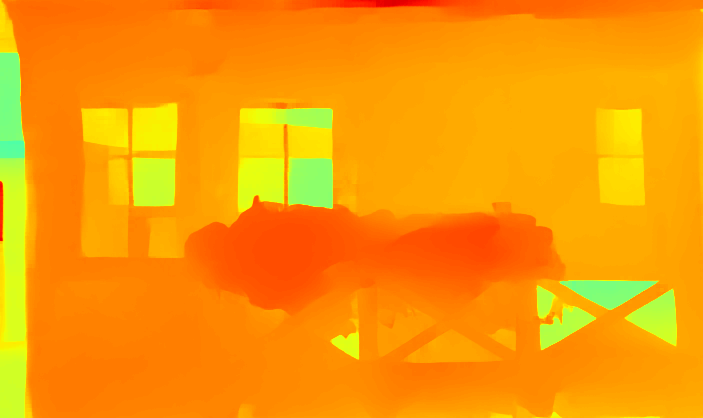} \hspace{-4mm} &
                \includegraphics[width=0.161\textwidth]{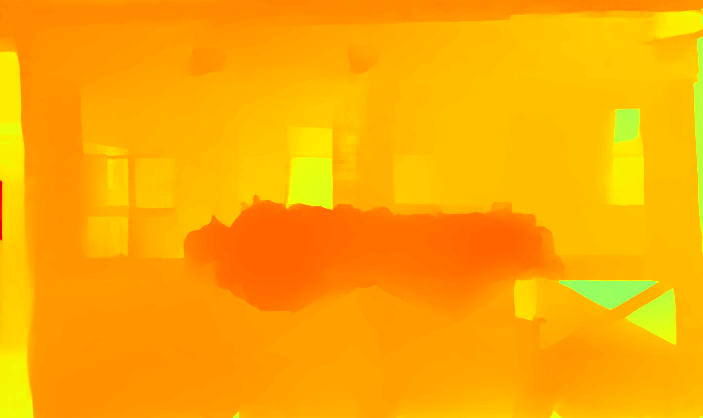} \hspace{-4mm} &
                \includegraphics[width=0.161\textwidth]{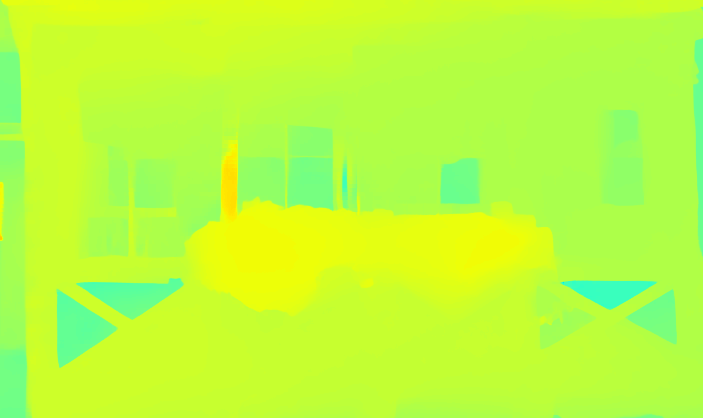} \hspace{-4mm}
                \\

                Bicubic \hspace{-4mm} &
                RealSR(CVPRW20')~\cite{ji2020real} \hspace{-4mm} &
                BSRGAN(ICCV21')~\cite{zhang2021designing} \hspace{-4mm} &
                SwinIR(ICCVW21')~\cite{liang2021swinir} \hspace{-4mm} &
                RealESRGAN(CVPR21')~\cite{wang2021real} \hspace{-4mm} &
                \\

                \includegraphics[width=0.161\textwidth]{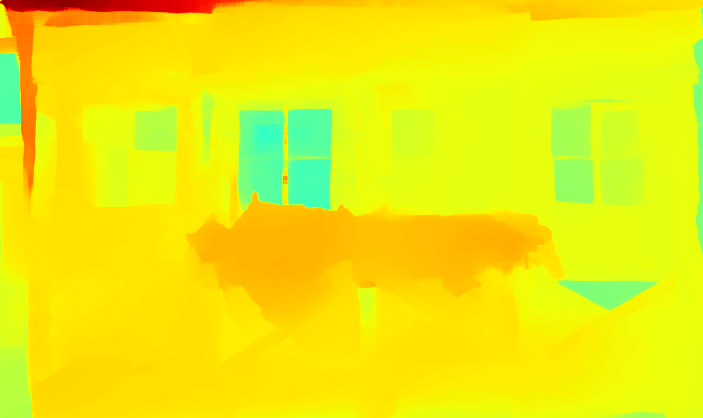} \hspace{-4mm} &
                \includegraphics[width=0.161\textwidth]{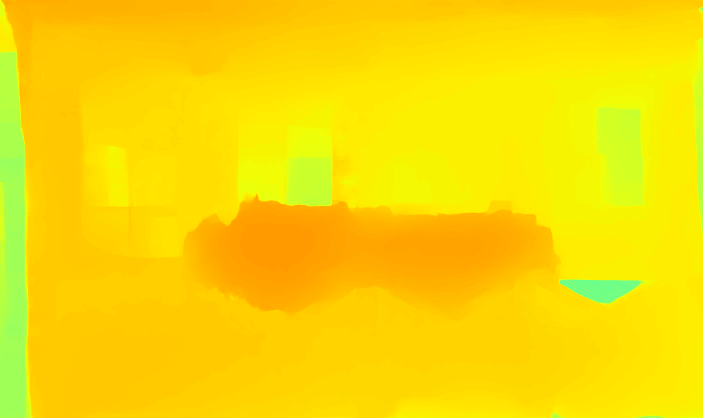} \hspace{-4mm} &
                \includegraphics[width=0.161\textwidth]{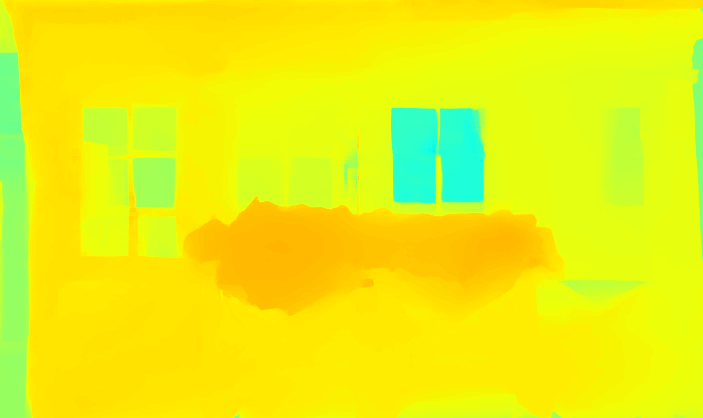} \hspace{-4mm} &
                \includegraphics[width=0.161\textwidth]{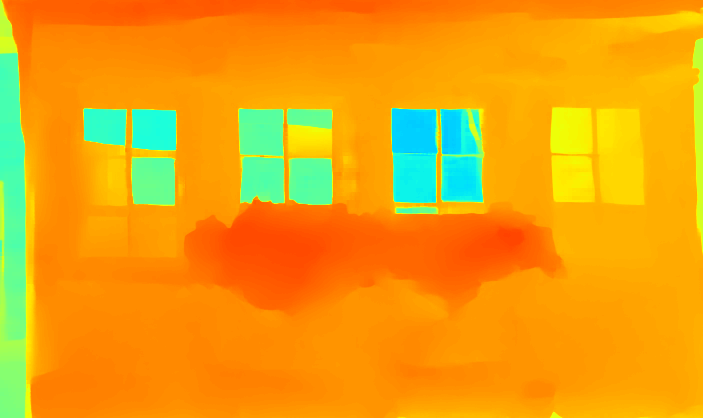} \hspace{-4mm}   &
                \includegraphics[width=0.161\textwidth]{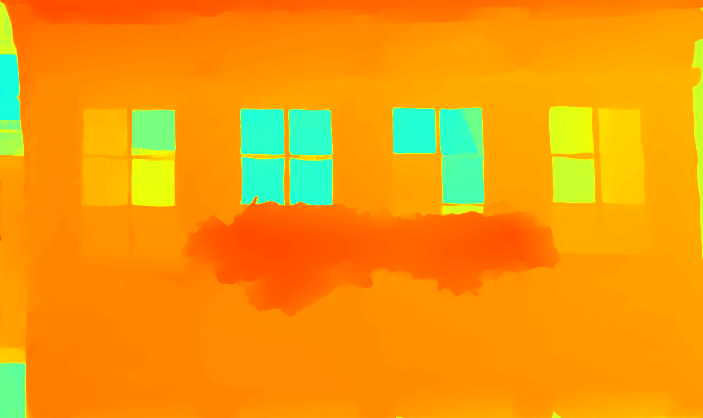} \hspace{-4mm}
                \\

                PDM-SRGAN(CVPR22')~\cite{luo2022learning} \hspace{-4mm} &
                DASR(ECCV22')~\cite{liang2022efficient}  \hspace{-4mm} &
                MMRealSR(ECCV22')~\cite{mou2022metric}  \hspace{-4mm} &
                RealSCGLAGAN(\textbf{Ours})    \hspace{-4mm} &
                GT \hspace{-4mm}

                \end{tabular}
                \end{adjustbox}
                \vspace{1mm}
                \\
         \end{tabular}}
		\captionof{figure}{Results of super-resolved and disparity estimated images obtained from several methods on the Flickr1024RS dataset, including RealSR \cite{ji2020real}, BSRGAN \cite{zhang2021designing}, RealESRGAN \cite{wang2021real}, SwinIR \cite{liang2021swinir}, PDM-SRGAN \cite{luo2022learning}, DASR \cite{liang2022efficient}, and MMRealSR \cite{mou2022metric}. }
    \label{fig:real_self_sr_results}
\end{center}
}] 
    \blfootnote{\IEEEauthorrefmark{1} These authors contributed equally to this work and should be considered co-first authors.}
    \blfootnote{Yuanbo Zhou, Graduate Student Member, IEEE. He is with the College of Physics and Information Engineering, Fuzhou University, Fuzhou, 350108, China. (e-mail: \url{webbozhou@gmail.com})}%
    \blfootnote{Yuyang Xue, Graduate Student Member, IEEE. He is of Engineering, University of Edinburgh (e-mail: \url{s2287251@ed.ac.uk})}%
    \blfootnote{Jiang Bi, is with the Beijing Radio and TV Station, Beijing, 100022, China. (e-mail: \url{bijinag1968@126.com})}
    \blfootnote{Wenlin He, is with the Beijing Radio and TV Station, Beijing, 100022, China. (e-mail: \url{hewenlin@sina.com})}
    \blfootnote{Xinlin Zhang, is with the College of Physics and Information Engineering, Fuzhou University, Fuzhou, 350108, China. (e-mail: \url{xinlin1219@gmail.com})}
    \blfootnote{Jiajun Zhang, is with the College of Physics and Information Engineering, Fuzhou University, Fuzhou, 350108, China. (e-mail: \url{211127180@fzu.edu.cn})}%
    \blfootnote{Wei Deng, is with the Imperial Vision Technology, Fuzhou, 350108, China. (e-mail: \url{weideng.chn@gmail.com})}
    \blfootnote{Nuofeng Nie, is with the Imperial Vision Technology, Fuzhou, 350108, China. (e-mail: \url{nieruofeng@imperial-vision.com})}
    \blfootnote{JunLin Lan, is with the College of Physics and Information Engineering, Fuzhou University, Fuzhou, 350108, China. (e-mail: \url{276215231@qq.com})}
    \blfootnote{Qinquan Gao, is with the College of Physics and Information Engineering, Fuzhou University, Fuzhou, 350108, China. (e-mail: \url{gqinquan@fzu.edu.cn})}
    \blfootnote{Tong Tong, is with the College of Physics and Information Engineering, Fuzhou University, Fuzhou, 350108, China. (email: \url{ttraveltong@gmail.com})}
\begin{abstract}

Real-world stereo image super-resolution has a significant influence on enhancing the performance of computer vision systems. Although existing methods for single-image super-resolution can be applied to improve stereo images, these methods often introduce notable modifications to the inherent disparity, resulting in a loss in the consistency of disparity between the original and the enhanced stereo images. To overcome this limitation, this paper proposes a novel approach that integrates a implicit stereo information discriminator and a hybrid degradation model. This combination ensures effective enhancement while preserving disparity consistency. The proposed method bridges the gap between the complex degradations in real-world stereo domain and the simpler degradations in real-world single-image super-resolution domain. Our results demonstrate impressive performance on synthetic and real datasets, enhancing visual perception while maintaining disparity consistency. The complete code is available at the following \href{https://github.com/fzuzyb/SCGLANet}{link}.

\end{abstract}


\section{Introduction}

Stereo vision is a technique widely employed for depth perception across multiple images and finds applications in various domains including robotics \cite{cosner2022self}, self-driving \cite{chuah2022semantic}, and AR/VR \cite{krajancich2020optimizing}, among others. Unfortunately, the accuracy and robustness of stereo image algorithms in real-world scenarios are often compromised by shooting noise and insufficient resolution. To address this issue, researchers have been investigating approaches for enhancing stereo images. In recent years, stereo image super-resolution has gained popularity due to its ability to improve image resolution while preserving natural and detailed textures.

The stereo image super-resolution model involves taking a pair of low-resolution images, denoted as $I_{L}^{LR}$ and $I_{R}^{LR}$, alongside their corresponding high-resolution images, denoted as $I_{L}^{HR}$ and $I_{R}^{HR}$, and learning a mapping function $F$ such that $F(I_{L}^{LR}, I_{R}^{LR}) \approx (I_{L}^{HR},I_{R}^{HR})$. Here, $I_{L}^{LR}$ and $I_{R}^{LR}$ represent the low-resolution left and right images, while $I_{L}^{HR}$ and $I_{R}^{HR}$ represent their respective high-resolution counterparts. The function $F$ is a complex nonlinear mapping typically represented by a deep learning model.

Considerable progress has been made in the field of stereo image super-resolution \cite{wang2019learning, ying2020stereo, song2020stereoscopic, wang2021symmetric, zhu2021cross, Chen2022Cross, chu2022nafssr}. However, most of these models rely on bicubic interpolation, which often introduces significant artifacts when applied to real-world stereo images. Alternatively, one can explore methods developed for real-world single-image super-resolution, such as \cite{wang2021real, he2021SRDRL, zhang2021designing} and \cite{mou2022metric}, to enhance stereo images. Nevertheless, when directly applied to stereo images, these methods tend to compromise the inherent disparities present in the original low-resolution stereo images.

\begin{figure}
\centering   
\includegraphics [width=0.4\textwidth]{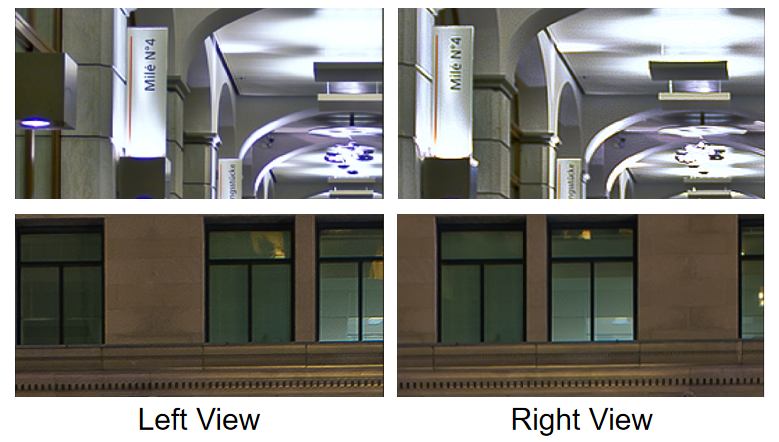}  
\caption{Images taken from the left and right perspectives of the same scene often exhibit varying levels of noise and blur ({\color[HTML]{0000FF} Zoom in for better visibility}).}
\label{fig_1} 
\end{figure}

Moreover, real-world scenes frequently experience variations in lighting conditions, focus, exposure time, and other factors throughout the capturing process. These variations introduce differences in the level of blurring and noise distribution between the left and right view images, as depicted in \cref{fig_1}. As a result, the degradation pattern in real-world stereo images is more complex than that in single images. This complexity significantly affects the training of stereo super-resolution models. Therefore, attempting to utilize single-image degradation models like BSRGAN \cite{zhang2021designing} or RealESRGAN \cite{wang2021real} to directly generate low-resolution data in the training process is not feasible.

Drawing inspiration from Chen et al.'s \cite{chen2017adversarial} approach in the domain of human pose estimation, which leverages prior information on human body structure in the discriminator to improve performance, we proposes a real-world stereo image super-resolution scheme that addresses the aforementioned challenges by utilizing the inherent prior information present in stereo images. Specifically, we introduce the implicit disparity prior information of the stereo images to the U-shape discriminator, compelling the generator to enhance stereo images while taking into account the disparity from the input low-resolution stereo images. Unlike a single-image super-resolution discriminator that only consists of a three-channel input layer, we design a six-channel input layer, with the first three channels representing the left view image and the last three channels representing the right view image. Both the left and right view images undergo feature extraction at different scales and are subsequently processed by the designed implicit disparity extraction module (IDEM), which reveals the feature-level disparity information contained within them.

To enhance the robustness of the trained stereo super-resolution model when dealing with complex degraded stereo images, we propose a stereo super-resolution generator that effectively integrates information from both the left and right images to restore severely degraded low-resolution stereo images. Additionally, a hybrid degradation model specifically tailored for stereo images is proposed to incorporates different noise kernels and blur kernels for the left and right views, addressing the issue of inconsistent blur and noise distribution. Furthermore, the shuffled degradation operation is also integrated to the second stage of the hybrid degradation model. This can enhance its ability to handle complex real-world stereo images.

The contributions of this paper are as follows:
\begin{itemize}
\item To the best of our knowledge, this paper is the first work that extends stereo super-resolution into the field of real-world scenes.
\item We present a stereo super-resolution training scheme and a discriminator that captures the inherent disparity of stereo images, forcing the generator to enhance stereo images while preserving the disparity information.
\item We propose a novel stereo-image hybrid degradation model that effectively processes more complex degraded stereo images.
\item Extensive experiments demonstrate that our proposed method achieves state-of-the-art performance on both synthetic and real datasets.
\end{itemize}

\section{Related Work}

\subsection{Single Image Super Resolution}

Convolutional neural networks have played a crucial role in advancing the field of single-image super-resolution over the past decade, as highlighted in the survey by Yang et al. \cite{Yang2019Deep}. The introduction of the SRCNN by Dong et al. (2015) marked a significant shift towards deep learning in super-resolution research. Presently, two primary research directions can be identified in single-image super-resolution: fidelity-based methods and perception-based methods. Fidelity-based methods, including VDSR \cite{kim2016accurate}, FSRCNN \cite{dong2016accelerating}, SRDenseNet \cite{tong2017image}, EDSR \cite{lim2017enhanced}, RCAN \cite{zhang2018image}, ISRN \cite{Liu2022Iterative} and HTI-Net~\cite{zhang2022heat}, employ L1 or L2 loss functions. On the other hand, perception-based approaches utilize generative adversarial loss \cite{goodfellow2020generative} and perceptual loss functions. Notable examples include SRGAN \cite{ledig2017photo}, ESRGAN \cite{wang2018esrgan}, RankSRGAN \cite{zhang2019ranksrgan} and FASRGAN \cite{Yan2021Fine}. Furthermore, the introduction of the transformer model \cite{vaswani2017attention} has paved the way for the utilization of vision-transformers \cite{dosovitskiy2020image} in super-resolution research. Examples include TTSR \cite{yang2020learning}, SwinIR \cite{liang2021swinir}, and ESRT \cite{lu2022transformer}.

Inspiring by these works, researchers made efforts on real-world super-resolution to address the issues of noise and blur associated with images in the real world. Yuan et al. \cite{yuan2018unsupervised} proposed CinCGAN, which considered unknown degradation and implemented a two-stage unsupervised approach to perform real-world super-resolution. In addition, Gu et al. proposed IKC \cite{gu2019blind}, which reduces artifacts when the model infers real-world images by estimating the blur kernel of the real world. Lugmayr et al. \cite{lugmayr2019unsupervised} proposed a different approach, using an unsupervised method to simulate and synthesize data in the real world to perform super-resolution. Subsequent works such as those proposed by Ji et al. \cite{ji2020real} designed a method to estimate blur kernels and noise, which achieved the best visual effect in the DPED dataset \cite{ignatov2017dslr}. However, since the degradation in real-world images is quite complex, Zhang et al. \cite{zhang2021designing} designed a simple and practical degradation model. By randomly combining different levels of blur, noise, compression artifacts, and varying scales of interpolation, the model employed paired images generated by the degradation model to train the super-resolution model, which improved the robustness of the model when tasked with handling complex degraded images. Recently, Wang et al. \cite{wang2021real} designed a high-order degradation model to further improve the performance of the super-resolution model on real-world images. Additionally, Mou et al. \cite{mou2022metric} proposed a metric learning technique to study inter-modulation in the real world, making it more adaptable to complex, real scenarios.

\subsection{Stereo Image Super Resolution}
The flourishing development of single-image super-resolution has provided a solid foundation for stereo image super-resolution. In stereo image super-resolution, not only can we optimize the model by using some of the techniques we use in single-image super-resolution, but we can also restore the texture of low-resolution images by using complementary information of stereo images. StereoSR \cite{jeon2018enhancing} horizontally shifts the image by different pixels and inputs them into the reconstruction network together, using left and right complementary information to enhance the performance of stereo super-resolution. Wang et al. \cite{wang2019learning} introduced a parallax attention mechanism, incorporating self-attention into the disparity attention module. Yan et al. \cite{yan2020disparity} proposed a new way to combine the feature modulation dense block (FMDB) with disparity attention loss, which effectively improves the performance of stereo super-resolution by using prior knowledge of disparity. Ying et al. \cite{ying2020stereo} introduced the stereo attention module (SAM), which can achieve stereo super-resolution by simply modifying the pre-trained single-image super-resolution model. Later, Song et al. \cite{song2020stereoscopic} further improved the performance of the stereo super-resolution model by combining self-attention and disparity attention mechanisms. Wang et al. \cite{wang2021symmetric} proposed the bilateral parallax attention module (biPAM), which uses the symmetry of stereo images to fuse cross-view data more effectively. In addition to CNN-based stereo super-resolution, recent researches have integrated vision transformers into stereo super-resolution. Kai et al. \cite{jin2022swinipassr} integrated Swin-Transformer \cite{liu2021swin} into stereo image super-resolution to improve the global perception ability of stereo super-resolution models. Meanwhile, Chu et al. \cite{chu2022nafssr} recently proposed the stereo cross-attention module (SCAM) to combine cross-view data and achieved the state-of-the-art performance on the Flickr1024 dataset \cite{wang2019flickr1024}.

Although the aforementioned stereo super-resolution methods have shown promising results in known degradation scenarios, their performance significantly deteriorates when applied to real-world stereo images, resulting in unsatisfactory artifacts. To the best of our knowledge, there is still no research dedicated to real-world stereo super-resolution. While it is possible to directly utilize similar methods used for real-world single-image super-resolution, such as BSRGAN \cite{zhang2021designing}, RealESRGAN \cite{wang2021real}, and MMRealSR \cite{mou2022metric}, we have observed that these methods heavily compromise the inherent disparity consistency of stereo images. Motivated by advancements in real-world single-image super-resolution and implicit knowledge embedding, this paper endeavors to address a series of challenging issues encountered by real-world stereo super-resolution models. 
\section{Proposed Method}

\begin{figure}[h]
    \centering
    \includegraphics [width=0.49\textwidth]{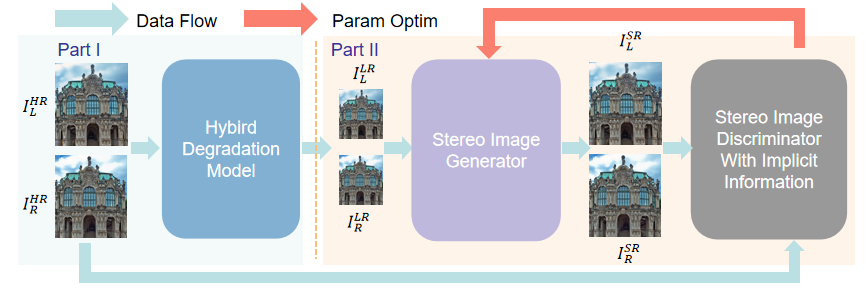}
    \caption{The framework of the proposed method consists of two parts: the hybrid degradation model and the stereo image super-resolution model.}
    \label{fig_2}
    \end{figure}

     In this section, we will provide a detailed explanation of the proposed method. \cref{fig_2} illustrates the overall flow of the method, comprising two main parts: the hybrid degradation model and the stereo super-resolution model. the hybrid degradation model will be introduced in \cref{sec:HDM} and the stereo super-resolution model will be presented in \cref{sec:SISRM}.

    \begin{figure*}[h]
    \centering
    \includegraphics [width=0.9\textwidth]{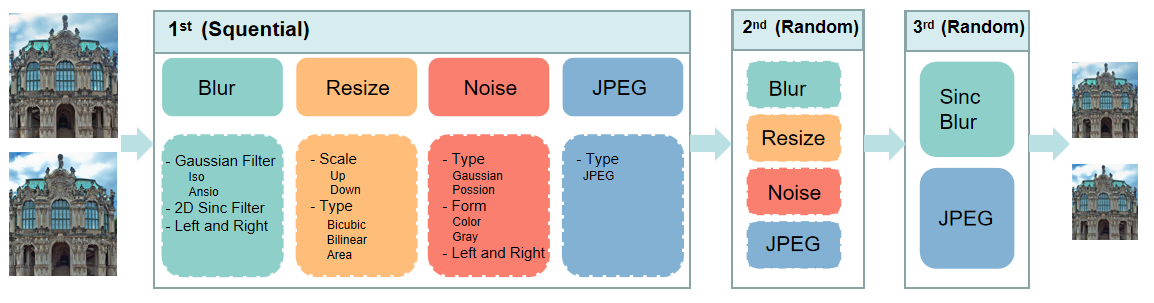}
    \caption{The hybrid degradation model of the proposed method.}
    \label{fig_3}

    \end{figure*}
    \begin{figure*}
    \centering
    \includegraphics [width=0.98\textwidth]{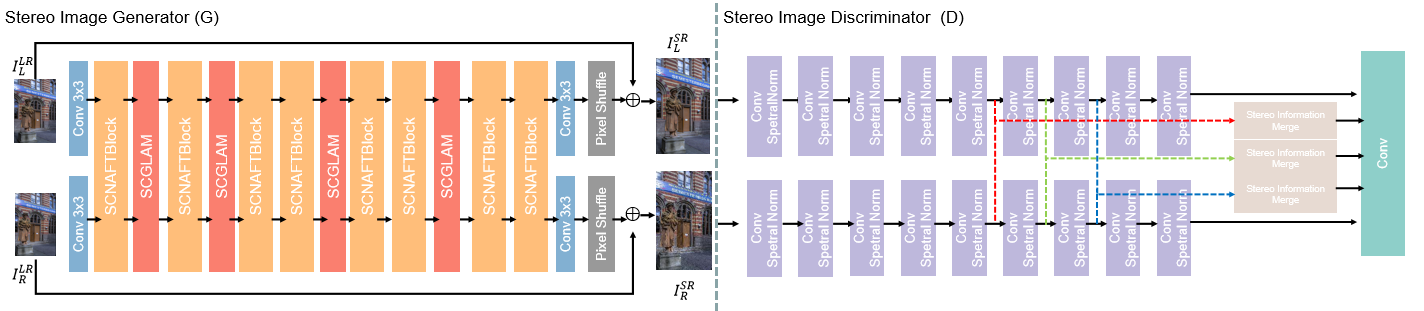}
    \caption{Figures of image super-resolution model, which consists of generator and discriminator.}
    \label{fig_4}
    \end{figure*}

\subsection{Hybrid Degradation Model}
    \label{sec:HDM}

   The proposed hybrid data degradation model consists of three stages, as shown in \cref{fig_3}.

   In the first stage, the degradation order remains fixed. The high-resolution image undergoes blur, up/downsampling, noise addition, and JPEG compression sequentially. This stage resembles traditional degradation models, but incorporates up-sampling operations to account for the possibility of up-sampling during the transmission of real images. The specific process can be described using \cref{equ:1}.

    \begin{equation}
    \label{equ:1}
        {I^{LR}} = T(I^{HR}) = [{({I^{HR}} * k)_ \updownarrow }_{_{\rm{s}}} + n)]_{\operatorname{JPEG}},
    \end{equation}


   \noindent where $I^{HR}=\{I_{L}^{HR}, I_{R}^{HR}\}$ represent the set of high-resolution images, $I_{L}^{HR}$ and $I_{R}^{HR}$ correspond to the left and right images, respectively. Similarly, $I^{LR}=\{I_{L}^{LR}, I_{R}^{LR}\}$ denotes the set of low-resolution images, with $I_{L}^{LR}$ and $I_{R}^{LR}$ representing the left and right images, respectively. The variable $k$ represents the blur kernel, which encompasses isotropic and anisotropic Gaussian filter kernels, as well as the Sinc filter kernel. The symbol $\updownarrow$ denotes the up/downsampling operator, utilizing bicubic, bilinear, and area methods, with $s$ representing the scale factor. The variable $n$ signifies noise, encompassing Gaussian and Poisson noise, while the term JPEG refers to the JPEG compression operator.

    In the second stage, a series of degradation operations are randomly shuffled to increase the variety within the degradation space. Specifically, the high-resolution image may undergo various degradation operations such as noise addition, JPEG compression, blur, and up/downsampling. Alternatively, it may undergo operations like JPEG compression, blur, up/downsampling, and noise addition. Furthermore, there is a probability of skipping certain degradation operations, resulting in only JPEG compression, noise addition, or other selected operations.

    Considering the noticeable differences in blur levels and noise distributions between left and right images in the real world (as shown in \cref{fig_1}), this paper intentionally introduces variations in noise levels and blur degrees between the left and right images during noise addition and blur degradation operations. This adjustment is formulated as \cref{equ:lr}.

    \begin{equation}
    \label{equ:lr}
    \begin{aligned}
    {I^{LR}_L} = T_l(I^{HR}_L)=[{({I^{HR}_L} * k_1)_ \updownarrow }_{_{\rm{s}}} + n_1)]_{\operatorname{JPEG}} \\
    {I^{LR}_R} = T_r(I^{HR}_R)=[{({I^{HR}_R} * k_2)_ \updownarrow }_{_{\rm{s}}} + n_2)]_{\operatorname{JPEG}},
    \end{aligned}
    \end{equation}
    \noindent where $k_1$ and $k_2$ represent different blur kernels, while $n_1$ and $n_2$ represent varying noise levels or types. This strategy expands the potential image degradations and improves the resilience of the trained stereo super-resolution model.

    In the third stage, we apply Sinc filtering, similar to RealESRGAN\cite{wang2021real}, to address the issues of ringing and overshoot artifacts. Additionally, to enhance the model's adaptation to a wider range of complex real-world data, this study integrates Sinc filtering with random JPEG compression, thereby increasing the diversity of the dataset.


\subsection{Stereo Image Super-resolution Model}
    \label{sec:SISRM}

    \cref{fig_4} illustrates the architecture of the stereo super-resolution model comprising the stereo image generator (SIG) and the stereo image discriminator (SID).

\begin{figure}[h]
    \centering
    \includegraphics [width=0.49\textwidth]{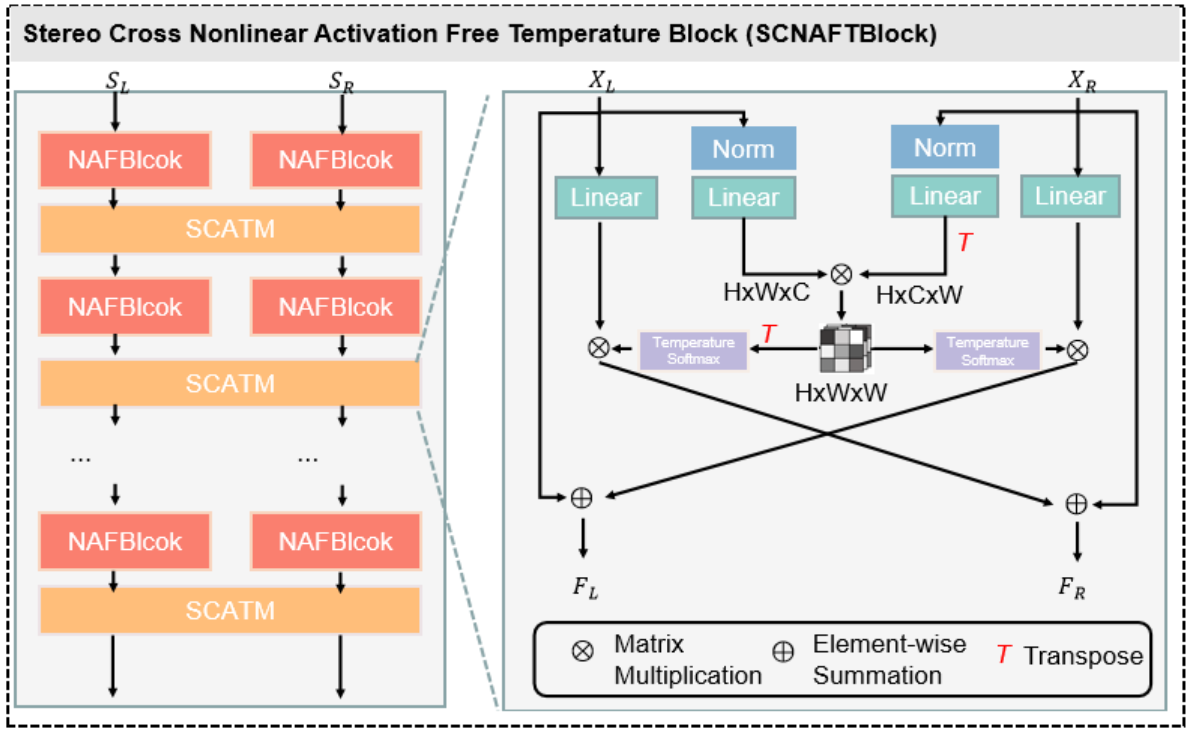}
    \caption{The Stereo Cross Nonlinear Activation Free Temperature Block (SCNAFTBlock), which consists of the NAFBlock \cite{chu2022nafssr} and the Stereo Cross Attention Temperature Module (SCATM) \cite{Zhou2023Stereo}.}
    \label{fig_SCNAFTBlock}
    \end{figure}

    \textbf{Stereo Image Generator (SIG).} The SIG is composed of multiple modules: a shallow feature extraction module, Stereo Cross Nonlinear Activation Free Temperature Block (SCNAFTBlock), Stereo Cross Global Learnable Attention Module (SCGLAM), and a reconstruction module. Specifically, the shallow feature extraction module consists of a 3x3 convolutional layer. Meanwhile, the SCNAFTBlock comprises the NAFBlock \cite{chu2022nafssr} and the Stereo Cross Attention Temperature module (SCATM) \cite{Zhou2023Stereo}. Additional details regarding the SCNAFTBlock can be found in \cref{fig_SCNAFTBlock}. On the other hand, the SCGLAM module effectively integrates similar textures from both intra-view and cross-view perspectives. Initially, the low-resolution left and right input images ($I_L^{LR}$ and $I_R^{LR}$) undergo shallow feature extraction through a shared 3x3 convolution module, resulting in shallow features ($S_L$ and $S_R$). These shallow features are further processed by the NAFBlock to obtain refined left and right features ($X_L$ and $X_R$), which are then inputted into the SCATM module to fuse information and generate $F_L$ and $F_R$. Subsequently, the SCGLAM module utilizes the fused $F_L$ and $F_R$ to capture intra-view and cross-view texture similarity, leading to $SF_L$ and $SF_R$. Finally, the reconstruction module, consisting of convolution and Pixshuffle \cite{shi2016real} modules, takes $SF_L$ and $SF_R$ as input to produce the reconstructed high-resolution left and right images ($I_L^{SR}$ and $I_R^{SR}$).

    In particular, the NAFBlock consists of a Mobile Convolution with channel attention and a feed-forward network (FFN). The computation process of the NAFBlock is illustrated in  \cref{euq:nafblock}.

    \begin{equation}
    \begin{aligned}
    &S=\operatorname{MBConv}(\operatorname{LN}(S))+ S \\
    &S=\operatorname{FFN}(\operatorname{LN}(S)+S
    \end{aligned}
    \label{euq:nafblock}
    \end{equation}

    \noindent where $S$ represents the input left/right features, LN represents LayerNorm~\cite{ba2016layer}. MBConv is composed of mobile convolutional layers~\cite{howard2019searching} and activation functions, while FFN consists of two layers of point convolution and activation functions. Notably, the activation function used in both MBConv and FFN is SimpleGate \cite{chu2022nafssr}, which operates as follows:

    \begin{equation}
    \operatorname{SimpleGate}(S)= S_1 \odot S_2,\label{eq:sg}
    \end{equation}

    \noindent where $S_1$ and $S_2$ are two equal feature maps obtained by dividing $S$ along the channel dimension, $\odot$ representing element-wise multiplication.

    For SCATM, its computation process can be represented by \cref{equ:scatm}.


    \begin{equation}
    \begin{aligned}
    & F_{R \rightarrow L}=\operatorname{TA}\left( W_1^{L} \overline{S}_{L}, W_1^{R} \overline{S}_{R}, W_2^{R} S_{R}\right), \\
    & F_{L \rightarrow R}=\operatorname{TA}\left( W_1^{R} \overline{S}_{R}, W_1^{L} \overline{S}_{L}, W_2^{L} S_{L}\right), \\
    & F_{L} = {\gamma}_{L} F_{R \rightarrow L} +  S_{L}, \\
    & F_{R} = {\gamma}_{R} F_{L \rightarrow R} +  S_{R}, \\
    \label{equ:scatm}
    \end{aligned}
    \end{equation}

    \noindent where  $\overline{S}_{L} = LN(S_{L})$, $\overline{S}_{R} = LN(S_{R})$. $W_1^{L}$, $W_1^{R}$, $W_2^{L}$ and $W_2^{R}$ are projection matrices. ${\gamma}_{L}$ and ${\gamma}_{R}$ are learnable scaling parameters. $\operatorname{TA}$ refers to the temperature attention module, which can be represented by \cref{equ:TA}.

     \begin{equation}
    \operatorname{TA}(Q, K, V)=\operatorname{softmax}\left( \tau Q K^T / \sqrt{C} \right) V,
    \label{equ:TA}
    \end{equation}

    \noindent where $\tau$ is a hyperparameter representing the temperature coefficient. By adjusting different temperature coefficients, the radio of fusion between left and right features can be controlled.


    In simple terms, the SCGLAM module functions as a global attention module within and between views. However, unlike traditional global attention, not all vectors require correlation computation. To reduce computational burden, we adopt Super-Bit Locality Sensitive Hashing (SB-LSH)~\cite{su2022global} to group input vectors into buckets, only performing correlations within each bucket. This approach effectively reduces computational complexity. Notably, we reshape the left and right feature maps, $F_{L}$ and $F_{R}$, into 1-D vectors and concatenate them to form the stereo feature space $F$. The bucketing process is defined by the following formula:

    \begin{equation}
    \lambda_i=\left\{x_j \mid argmax \left(Mx_i \right)= argmax \left(Mx_j\right)\right\},
    \label{equ:labmda}
    \end{equation}

    \noindent where $M \in R^{b \times c}$ is an orthogonal matrix, $b$  represents the number of hash buckets, $x_{j}$ and $x_{k}$ are the $j$-th and $k$-th feature vectors in $F \in R^{c \times 2hw}$, and $\lambda_i$ represents the index set of the bucket. After bucketing is completed, correlation calculation can be performed, which can be represented by \cref{equ:scgla}.

    \begin{equation}
    SCGLA\left(x_i\right)=\sum_{x_j \in \lambda_i} \frac{\exp \left(s\left(x_i, x_j\right)\right)}{\sum_{x_k \in \lambda_i} \exp \left(s\left(x_i, x_k\right)\right)} \phi_v\left(x_j\right),
    \label{equ:scgla}
    \end{equation}

    \noindent where $\phi_v(\cdot)$ is a feature embedding layer, $s(\cdot,\cdot)$ is used to calculate the similarity between two vectors. It consists of a learnable scoring function and a fixed dot product similarity function. Specifically, it can be represented as \cref{equ:s}.

     \begin{equation}
        s(x_i,x_j) = s^{l}_{j}(x_i) + s_f(x_i,x_j),
        \label{equ:s}
    \end{equation}

    \noindent where $s_f(x_i,x_j) = \phi_q\left(x_i\right)^{\mathrm{T}} \phi_k\left(x_j\right)$, $\phi_q(\cdot)$ and $\phi_k(\cdot)$ are shared feature embedding layers, $s^{l}_{j}(x_i)$ is the $j$-th component in $s_l(x_i)$, which can be defined as \cref{equ:sl}. The specific implementation details of SCGLAM can be found in supplementary material.

    \begin{equation}
    s_l(x_i) = W_2 \sigma\left(W_1\phi_l\left(x_i\right)+b_1\right)+b_2,
    \label{equ:sl}
    \end{equation}

    \noindent where $\sigma(\cdot)$ is the ReLU activation and $W_1$,$W_2$,$b_1$,$b_2$ are learnable parameters.

    \textbf{Stereo Image Discriminator (SID).}  The SID in stereo image super-resolution has a distinct role compared to its equivalent in single image super-resolution. In addition to discriminating visual perceptual differences between reconstructed and ground truth (GT) images, it also handles the differences of disparity between the reconstructed and GT images. Consequently, the development of an effective SID for stereo image super-resolution is a challenging task. One possible solution is to augment the discriminator with a branch to estimate the disparity of the left and right views. However, the existing stereo disparity estimation models are either too large or lack the necessary accuracy, making this approach less feasible.

\begin{figure}
    \centering
    \includegraphics [width=0.45\textwidth]{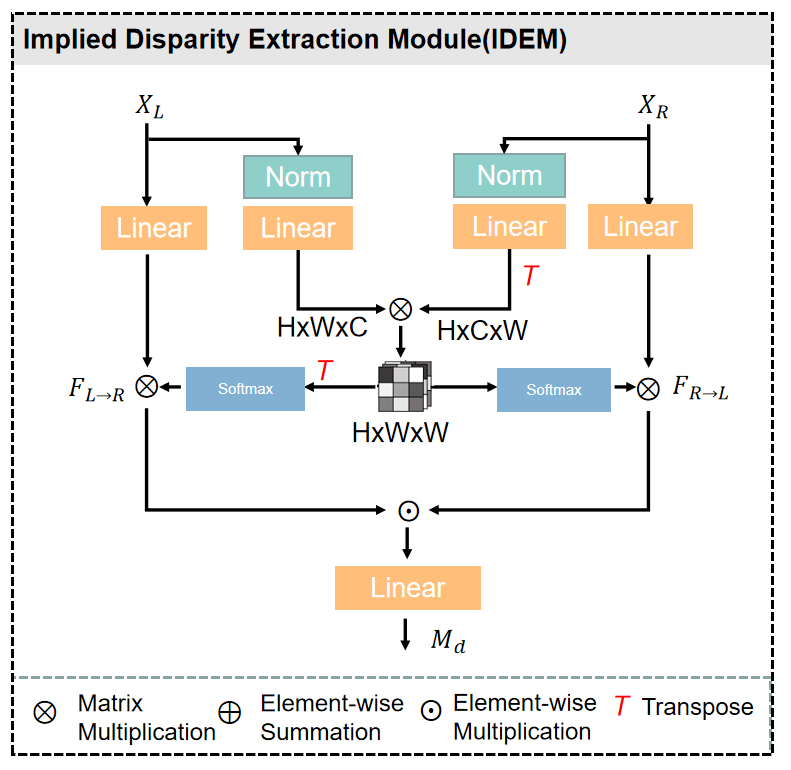}
    \caption{The implicit disparity extraction module (IDEM).}
    \label{fig_idem}
    \end{figure}

    Inspired by the field of human pose estimation \cite{chen2017adversarial}, researchers have incorporated the structural information of the human body into the discriminator to enhance accuracy. In line with this, our paper integrates disparity information implicitly into the discriminator. Specifically, we design a composed discriminator that can effectively capture both the visual perceptual differences of images and the implicit disparity information. This composed discriminator consists of a U-shaped network with shared weights, several implicit disparity extraction modules (IDEMs), and a fusion convolution. To provide comprehensive description, \cref{fig_4} illustrates the composition of this discriminator. Given left and right images $I_L^{SR}$ and $I_R^{SR}$, the U-shaped network generates multi-scale features $F_{\operatorname{x}1}, F_{\operatorname{x}2}$ and $F_{\operatorname{x}4} $. These multi-scale features are then processed by the implicit disparity extraction module, producing implicit disparity information at different scales. Additionally, to capture the specific visual perceptual information of the left and right views, the features of these views are directly fed into the last fusion convolution layer, bypassing the implicit disparity extraction module. This fusion layer not only integrates implicit disparity information from multiple scales but also incorporates image-specific visual perceptual information. Consequently, the discriminator is capable of effectively discerning the visual perceptual information unique to the left and right images, as well as the implicit disparity information. The complete process of the discriminator is formulated in \cref{equ:dis}.

    \begin{equation}
    \begin{aligned}
    &F_{\operatorname{x}1},F_{\operatorname{x}2},F_{\operatorname{x}4} = \operatorname{UNet}([I_{L}^{SR},I_{R}^{SR}]),  \\
    &M_{d\operatorname{x}1},M_{d\operatorname{x}2},M_{d\operatorname{x}4} = \operatorname{IDEM}(F_{\operatorname{x1}},F_{\operatorname{x2}},F_{\operatorname{x4}}), \\
    &D_o = \operatorname{Conv}(M_{d\operatorname{x}1},M_{d\operatorname{x}2},M_{d\operatorname{x}4},\operatorname{ConvSN}(F_{\operatorname{x}1})), \\
    \label{equ:dis}
    \end{aligned}
    \end{equation}

    \noindent where $M_{d\operatorname{x}1},M_{d\operatorname{x}2}$ and $M_{d\operatorname{x}4}$ represent the implicit disparity information at x1, x2, and x4 scales, respectively. ConvSN is composed of a convolution layer, ReLU activation function, and Spectral Normalization \cite{miyato2018spectral}. It is important to note that in the final fusion convolution process, $M_{d\operatorname{x}1}, M_{d\operatorname{x}2}, M_{d\operatorname{x}4}$, and ConvSN($F_{\operatorname{x1}}$) are interpolated to the same spatial resolution for effective information fusion. The Implicit Disparity Extraction Module (IDEM), as illustrated in \cref{fig_idem}, is responsible for the extraction of implicit disparity. The functional expression can be described by \cref{equ:2}.

    \begin{equation}
    \begin{aligned}
    & F_{R \rightarrow L}=\operatorname{Attention}\left(W_1^{L} \overline{X}_{L}, W_1^{R} \overline{X}_{R}, W_2^{R} X_{R}\right), \\
    & F_{L \rightarrow R}=\operatorname{Attention}\left(W_1^{R} \overline{X}_{R}, W_1^{L} \overline{X}_{L}, W_2^{L} X_{L}\right), \\
    &F_{d} = F_{R \rightarrow L} \odot F_{L \rightarrow R}, \\
    &M_{d} = W_{c}F_{d}, \\
    \label{equ:2}
    \end{aligned}
    \end{equation}

    \noindent  where $X_{L}$  and $X_{R}$ represent the input left and right feature maps, respectively. $\overline{X}_{L} = LN( X_{L})$, $\overline{X}_{R} = LN( X_{R})$.
    Attention can be described as \cref{equ:3}.
    \begin{equation}
    \operatorname{Attention}(Q, K, V)=\operatorname{softmax}\left(Q K^T / \sqrt{C}\right) V,
    \label{equ:3}
    \end{equation}

    \noindent where, $C$ represents the dimension of the feature.

\subsection{Loss Function}
    In this section, the loss function that was used in detail will be introduced. Similar to single-image super-resolution, the loss function for the generator also consists of three major parts: pixel loss, perceptual loss, and adversarial loss. The pixel loss is represented by L1 and can be written as \cref{equ:4}.

    \begin{equation}
    \mathcal{L}_{pixel}=\left\|I^{SR}-I^{HR}\right\|_1,
    \label{equ:4}
    \end{equation}

    \noindent where $I^{HR}=\{I_{L}^{HR}, I_{R}^{HR}\}$ represents the GT stereo images, $I^{SR}=\{I_{L}^{SR}, I_{R}^{SR}\}$ represents the super-resolved stereo images.


    Inspired by Ma et al. \cite{ma2021perception}, which employed residual information to train model. We combined the residual information of the right image (obtained by subtracting the left image) to address the limitation of the VGG model \cite{simonyan2014very} in incorporating the implicit information from the left and right images. In the end, the perceptual loss is calculated by \cref{equ:5}.

    \begin{equation}
    \begin{split}
    \mathcal{L}_{\text{per}}= &\text{LPIPS}(I_{L}^{SR}-I_{L}^{HR})\\
                       &+\text{LPIPS}(I_{R}^{SR}-I_{R}^{HR}) \\
                       &+\varepsilon \text{LPIPS}(Res^{SR}-Res^{HR}),
    \label{equ:5}
    \end{split}
    \end{equation}

    \noindent where $Res$ denotes the residual value between the left and right images, namely, $Res^{SR}=I_{R}^{SR}-I_{L}^{SR}$ and $Res^{HR}=I_{R}^{HR}-I_{L}^{HR}$. LPIPS represents the LPIPS metric, which can be found in \cite{zhang2018unreasonable}. Further, $\varepsilon$ represents the perceptual weight of the residual component. The adversarial loss, which can be formulated as \cref{equ:6}.

    \begin{equation}
    \mathcal{L_G}= \mathbb{E}_{I^{SR}}[1-D(I^{SR})],
    \label{equ:6}
    \end{equation}

    \noindent where $D$ denotes the discriminator. Therefore, the entire loss of the generator can be written as \cref{equ:7}.

    \begin{equation}
    \mathcal{L}_{Gtotal}= \gamma \mathcal{L_{\text {per }}}+ \lambda \mathcal{L_G} + \eta \mathcal{L}_{pixel},
    \label{equ:7}
    \end{equation}

    \noindent where $\gamma$, $\lambda $ and $\eta$ denote the weight of different loss.

    The loss of the discriminator can be written as \cref{equ:8}.

    \begin{equation}
    \mathcal{L_D}= \mathbb{E}_{I^{SR}}[-D(I^{SR})].
    \label{equ:8}
    \end{equation} 
\section{Experiments And Analysis}

\subsection{Dataset}
\textbf{Training dataset.} In this research, to ensure a fair comparison with other state-of-the-art methods, similar to NAFSSR \cite{chu2022nafssr}, the training dataset comprises 800 pairs from the Flickr1024 dataset \cite{wang2019flickr1024} and 60 pairs from the Middlebury dataset \cite{scharstein2014high}.

\begin{figure}
        \centering
        \includegraphics [width=0.45\textwidth]{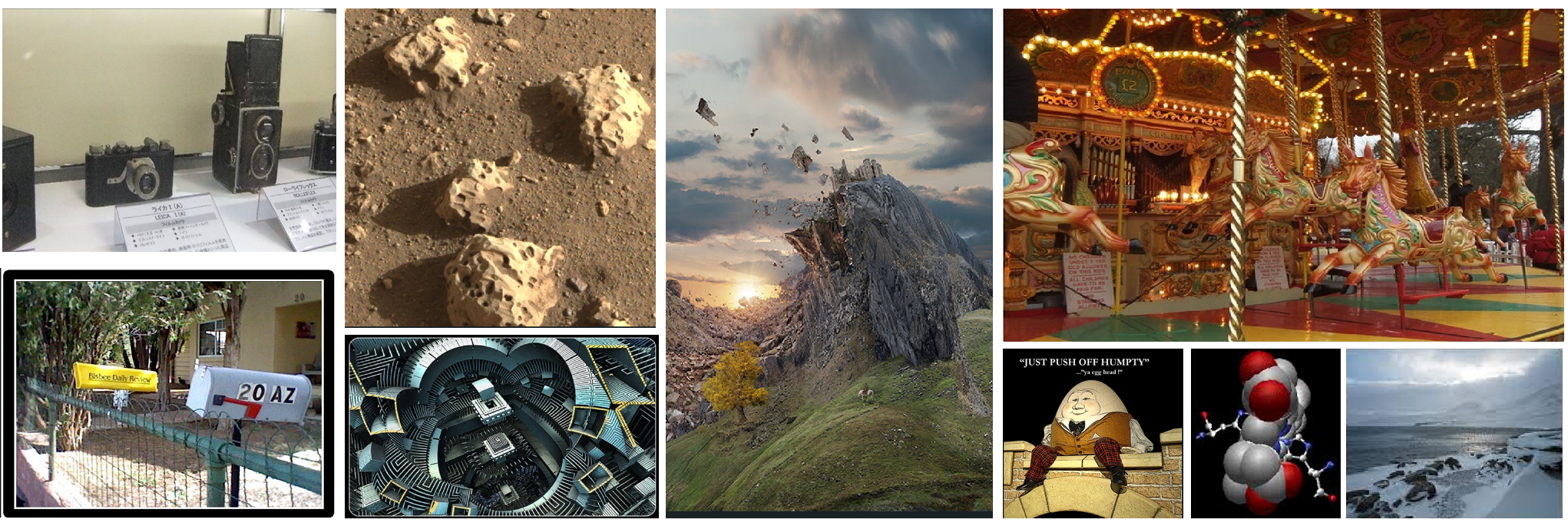}
        \caption{Some cases of StereoWeb20.}
        \label{fig_6}
    \end{figure}

\textbf{Test Dataset.} To evaluated the model's fidelity, four commonly datasets were used, including KITTI 2012 \cite{geiger2012we}, KITTI 2015 \cite{geiger2015kitti}, Middlebury \cite{scharstein2014high}, and Flickr1024 \cite{wang2019flickr1024}. The KITTI 2012 dataset consists of 20 pairs of images, the KITTI 2015 dataset consists of 20 pairs of images, the Middlebury dataset consists of 5 pairs of images, and the Flickr1024 dataset consists of 112 pairs of images. To assess the proposed method's performance on real-world datasets, we created a new test dataset called Flickr1024RS by degrading the 112 pairs of images from Flickr1024. The degradation parameters used were as follows: blur sigma=[0.2,1.5], resize range=[0.5,1.2], noise range=[1,15], and jpeg range=[30,95]. Additionally, we obtained 20 pairs of real low-resolution stereo images from the internet, which were preprocessed and cropped for no-reference image quality assessment. We named this test set StereoWeb20, and some sample images from StereoWeb20 are shown in \cref{fig_6}. To facilitate future comparisons, the Flickr1024RS and StereoWeb20 test datasets can be directly downloaded from this \href{https://github.com/fzuzyb/SCGLANet}{link}.

\begin{table*}[!htb]
\centering
\caption{Quantitative results achieved by different methods on the KITTI 2012 \cite{geiger2012we}, KITTI 2015 \cite{geiger2015kitti}, Middlebury \cite{scharstein2014high}, and Flickr1024 \cite{wang2019flickr1024} datasets.
} \label{tab:flickr1024_kitti_mid}

\resizebox{\textwidth}{!}
{
\begin{tabular}{lccccccccc}
\toprule
\multirow{2}*{Methods} & \multirow{2}*{Scale} & \multirow{2}*{$P$$^{\dag}$} & \multicolumn{3}{c}{\textit{Left}$^{*}$ }& \multicolumn{4}{c}{$\left(\textit{Left}+\textit{Right} \right)/2$$^{*}$ }\\
\cmidrule(lr){4-6} \cmidrule(lr){7-10}
         &      &           & KITTI 2012 & KITTI 2015 & Middlebury & KITTI 2012 & KITTI 2015 & Middlebury & Flickr1024\\
\midrule
VDSR~\cite{kim2016accurate} & $\times$2 & 0.66M & 30.17$/$0.9062 & 28.99$/$0.9038 & 32.66$/$0.9101 & 30.30$/$0.9089 & 29.78$/$0.9150& 32.77$/$0.9102 & 25.60$/$0.8534\\
EDSR~\cite{lim2017enhanced} & $\times$2 & 38.6M & 30.83$/$0.9199 & 29.94$/$0.9231 & 34.84$/${0.9489} &30.96$/$0.9228 & 30.73$/$0.9335 & {34.95}$/${0.9492} & {28.66}$/$0.9087 \\
RDN~\cite{zhang2018residual} & $\times$2 & 22.0M  & 30.81$/$0.9197 & 29.91$/$0.9224 & {34.85}$/$0.9488 &30.94$/$0.9227 & 30.70$/$0.9330 & 34.94$/$0.9491 & 28.64$/$0.9084 \\
RCAN~\cite{zhang2018image} & $\times$2 & 15.3M & 30.88$/$0.9202 & 29.97$/$0.9231 & 34.80$/$0.9482 & 31.02$/$0.9232 & 30.77$/$0.9336 & 34.90$/$0.9486 & 28.63$/$0.9082 \\
StereoSR~\cite{jeon2018enhancing} & $\times$2 &1.08M & 29.42$/$0.9040 & 28.53$/$0.9038 & 33.15$/$0.9343 & 29.51$/$0.9073 & 29.33$/$0.9168 & 33.23$/$0.9348 & 25.96$/$0.8599 \\
PASSRnet~\cite{wang2019learning} & $\times$2 & 1.37M & 30.68$/$0.9159 & 29.81$/$0.9191 & 34.13$/$0.9421 & 30.81$/$0.9190 & 30.60$/$0.9300 & 34.23$/$0.9422 & 28.38$/$0.9038 \\
IMSSRnet~\cite{lei2020deep} & $\times$2 & 6.84M & 30.90$/$- & 29.97$/$- & 34.66$/$- & 30.92$/$- & 30.66$/$- & 34.67$/$- & -$/$- \\
iPASSR~\cite{wang2021symmetric} & $\times$2 & 1.37M & {30.97}$/${0.9210} & {30.01}$/${0.9234} & 34.41$/$0.9454 & {31.11}$/${0.9240} & {30.81}$/${0.9340} & 34.51$/$0.9454 & 28.60$/${0.9097} \\
SSRDE-FNet~\cite{dai2021feedback}   & $\times$2 & 2.10M & {31.08}$/${0.9224} & {30.10}$/${0.9245} & {35.02}$/${0.9508} & {31.23}$/${0.9254} & {30.90}$/${0.9352} & {35.09}$/${0.9511} & {28.85}$/${0.9132} \\

NAFSSR-B \cite{chu2022nafssr} & $\times$2 & 6.77M  & 31.40$/$0.9254 & 30.42$/${0.9282} & {35.62}$/${0.9545} & {31.55}$/${0.9283} & {31.22}$/${0.9380} & {35.68}$/${0.9544} & {29.54}$/${0.9204} \\

NAFSSR-L \cite{chu2022nafssr} & $\times$2 & 23.79M  & \underline{31.45}$/$\underline{0.9261} & \underline{30.46}$/$\underline{0.9289} & \underline{35.83}$/$\underline{0.9559} & \underline{31.60}$/$\underline{0.9291} & \underline{31.25}$/$\underline{0.9386} & \underline{35.88}$/$\underline{0.9557} & \underline{29.68}$/$\underline{0.9221} \\

SCGLANet (\textbf{Ours}) & $\times$2 & 25.43M  & \textbf{31.61}$/$\textbf{0.9294} & \textbf{31.23}$/$\textbf{0.9385} & \textbf{36.26}$/$\textbf{0.9587} & \textbf{31.71}$/$\textbf{0.9307} & \textbf{31.32}$/$\textbf{0.9397} & \textbf{36.33}$/$\textbf{0.9583} & \textbf{29.97}$/$\textbf{0.9257} \\
\midrule
\midrule
VDSR~\cite{kim2016accurate} &  $\times$4 & 0.66M & 25.54$/$0.7662 & 24.68$/$0.7456 & 27.60$/$0.7933 & 25.60$/$0.7722 & 25.32$/$0.7703 & 27.69$/$0.7941 & 22.46$/$0.6718 \\
EDSR~\cite{lim2017enhanced} &  $\times$4 & 38.9M & 26.26$/$0.7954 & 25.38$/$0.7811 & 29.15$/${0.8383} & 26.35$/$0.8015 & 26.04$/$0.8039 & 29.23$/$0.8397 & 23.46$/$0.7285 \\
RDN~\cite{zhang2018residual} &  $\times$4 & 22.0M  & 26.23$/$0.7952 & 25.37$/$0.7813 & 29.15$/$0.8387 & 26.32$/$0.8014 & 26.04$/$0.8043 & 29.27$/${0.8404} & 23.47$/${0.7295} \\
RCAN~\cite{zhang2018image} &  $\times$4 & 15.4M & 26.36$/$0.7968 & 25.53$/$0.7836 & {29.20}$/$0.8381 & 26.44$/$0.8029 & 26.22$/$0.8068 & {29.30}$/$0.8397 & {23.48}$/$0.7286 \\
StereoSR~\cite{jeon2018enhancing}  &  $\times$4 & 1.42M   & 24.49$/$0.7502 & 23.67$/$0.7273 &27.70$/$0.8036 & 24.53$/$0.7555 & 24.21$/$0.7511 & 27.64$/$0.8022 & 21.70$/$0.6460 \\
PASSRnet~\cite{wang2019learning}  &  $\times$4 & 1.42M   & 26.26$/$0.7919 & 25.41$/$0.7772 &28.61$/$0.8232 & 26.34$/$0.7981 & 26.08$/$0.8002 & 28.72$/$0.8236 & 23.31$/$0.7195 \\
SRRes+SAM~\cite{ying2020stereo}  &  $\times$4 & 1.73M  & 26.35$/$0.7957 & 25.55$/$0.7825 & 28.76$/$0.8287 & 26.44$/$0.8018 & 26.22$/$0.8054 & 28.83$/$0.8290 & 23.27$/$0.7233 \\
IMSSRnet~\cite{lei2020deep} &  $\times$4 & 6.89M  & 26.44$/$- & 25.59$/$- & 29.02$/$- & 26.43$/$- & 26.20$/$- & 29.02$/$- & -$/$- \\
iPASSR~\cite{wang2021symmetric}  &  $\times$4 & 1.42M  & {26.47}$/${0.7993} & {25.61}$/${0.7850} & 29.07$/$0.8363 & {26.56}$/${0.8053} & {26.32}$/${0.8084} & 29.16$/$0.8367 & 23.44$/$0.7287 \\
SSRDE-FNet~\cite{dai2021feedback}  & $\times$4 & 2.24M  & {26.61}$/${0.8028} & {25.74}$/${0.7884} & {29.29}$/${0.8407} & {26.70}$/${0.8082} & {26.43}$/${0.8118} & {29.38}$/${0.8411} & {23.59}$/${0.7352} \\

NAFSSR-B \cite{chu2022nafssr} & $\times$4 & 6.80M  & {26.99}$/${0.8121} & {26.17}$/${0.8020} & {29.94}$/${0.8561} & {27.08}$/${0.8181} & {26.91}$/${0.8245} & {30.04}$/${0.8568} & {24.07}$/${0.7551} \\
NAFSSR-L \cite{chu2022nafssr} & $\times$4 & 23.83M  & \underline{27.04}$/$\underline{0.8135} & \underline{26.22}$/$\underline{0.8034} & \underline{30.11}$/$\underline{0.8601} & \underline{27.12}$/$\underline{0.8194} & \underline{26.96}$/$\underline{0.8257} & \underline{30.20}$/$\underline{0.8605} & \underline{24.17}$/$\underline{0.7589} \\

SCGLANet (\textbf{Ours}) & $\times$4 & 25.29M  & \textbf{27.10}$/$\textbf{0.8204} & \textbf{26.94}$/$\textbf{0.8268} & \textbf{30.23}$/$\textbf{0.8628} & \textbf{27.20}$/$\textbf{0.8230} & \textbf{27.01}$/$\textbf{0.8290} & \textbf{30.38}$/$\textbf{0.8642} & \textbf{24.39}$/$\textbf{0.7679} \\

\bottomrule
\multicolumn{10}{l}{$\dag$ $P$ represents the number of parameters of the networks.} \\
\multicolumn{10}{l}{$*$ The best results are in \textbf{bold faces} and the second results are in \underline{underline}.}\\
\end{tabular}}
\vspace{-3mm}
\end{table*}

\subsection{Evaluation Metrics}
The performance of classic stereo super-resolution was evaluated using the peak signal-to-noise ratio (PSNR) and structural similarity index measure (SSIM). Additionally, for the evaluation of real-world stereo image super-resolution, the LPIPS \cite{zhang2018unreasonable} and Mean Absolute Disparity Error (MADE) metrics were employed. MADE measures the average absolute disparity error between the enhanced stereo images and the ground truth (GT) stereo images. The disparity calculation can be referred to as \cref{equ:made}, following the methodology described in \cite{Wang2023NTIRE}. Since StereoWeb20 lacks ground truth, the no-reference image quality evaluation metrics NRQM \cite{ma2017learning} and PI \cite{blau20182018} were adopted.

\begin{equation}
        \text{MADE} = \text{MAE}\left(Dis(I^{SR}), Dis(I^{HR})\right),
\label{equ:made}
\end{equation}

\noindent where $Dis(\cdot)$ is the disparity function and can be calculated using \cite{lipson2021raft}.


\subsection{Training Details}

The training process consists of two stages, following a similar approach as RealESRGAN \cite{wang2021real}. Initially, the generator was trained using only the pixel loss, followed by the incorporation of perception loss and adversial loss. The first and second stages involved training for 300,000 and 100,000 iterations, respectively, utilizing four RTX3090s. The training process employed the Adam optimizer \cite{kingma2014adam} and cosine annealing strategy \cite{loshchilovstochastic}, starting with an initial learning rate of 1e-4 and gradually decreasing to a final learning rate of 1e-7. A batch size of 12 was used. In the second stage, the loss function was parameterized with $\gamma$:$\lambda$:$\eta$:$\varepsilon$ = 1:0.1:1:0.1, and a patch size of 96 $\times$ 304 was utilized.

\begin{figure*}[htb]
\scriptsize
\centering
\begin{tabular}{cc}
\hspace{-0.4cm}
\begin{adjustbox}{valign=t}
\begin{tabular}{c}
\includegraphics[width=0.12\textwidth,height=0.149\textheight]{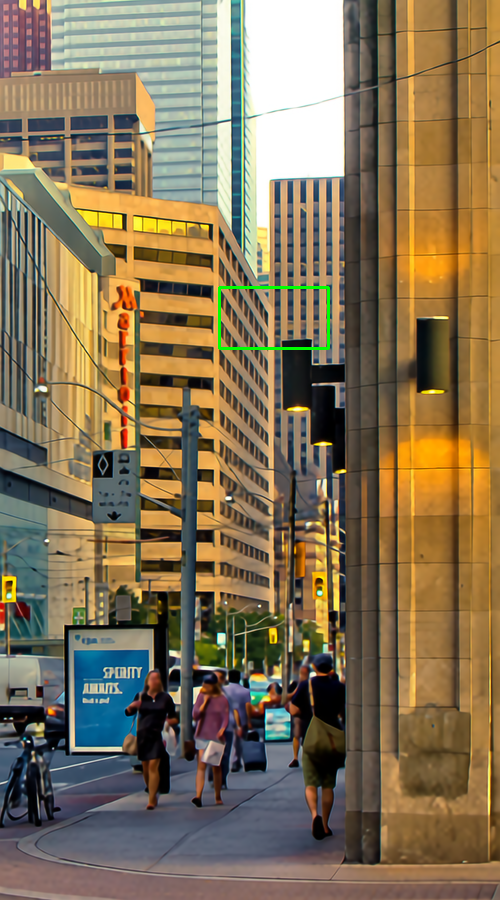}
\\
img\_0023
\end{tabular}
\end{adjustbox}
\hspace{-0.46cm}
\begin{adjustbox}{valign=t}
\begin{tabular}{cccccc}
\includegraphics[width=0.161\textwidth]{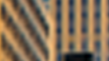} \hspace{-4mm} &
\includegraphics[width=0.161\textwidth]{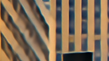} \hspace{-4mm} &
\includegraphics[width=0.161\textwidth]{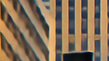} \hspace{-4mm} &
\includegraphics[width=0.161\textwidth]{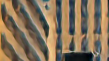} \hspace{-4mm} &
\includegraphics[width=0.161\textwidth]{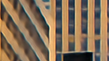} \hspace{-4mm}
\\

Bicubic \hspace{-4mm} &
RDN~\cite{zhang2018residual} \hspace{-4mm} &
RCAN~\cite{zhang2018image} \hspace{-4mm} &
StereoSR~\cite{jeon2018enhancing} \hspace{-4mm} &
SRRes+SAM~\cite{ying2020stereo} \hspace{-4mm} &
\\

\includegraphics[width=0.161\textwidth]{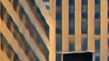} \hspace{-4mm} &
\includegraphics[width=0.161\textwidth]{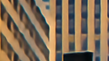} \hspace{-4mm} &
\includegraphics[width=0.161\textwidth]{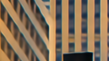} \hspace{-4mm} &
\includegraphics[width=0.161\textwidth]{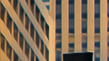} \hspace{-4mm}   &
\includegraphics[width=0.161\textwidth]{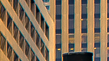} \hspace{-4mm}
\\

iPASSR~\cite{wang2021symmetric} \hspace{-4mm} &
SSRDE-FNet~\cite{dai2021feedback}  \hspace{-4mm} &
NAFSSR-L ~\cite{chu2022nafssr}  \hspace{-4mm} &
SCGLANet (\textbf{Ours})  \hspace{-4mm} &
Ground Truth \hspace{-4mm}
\\
\end{tabular}
\end{adjustbox}
\vspace{1mm}
\\

\hspace{-0.4cm}
\begin{adjustbox}{valign=t}
\begin{tabular}{c}
\includegraphics[width=0.12\textwidth,height=0.173\textheight]{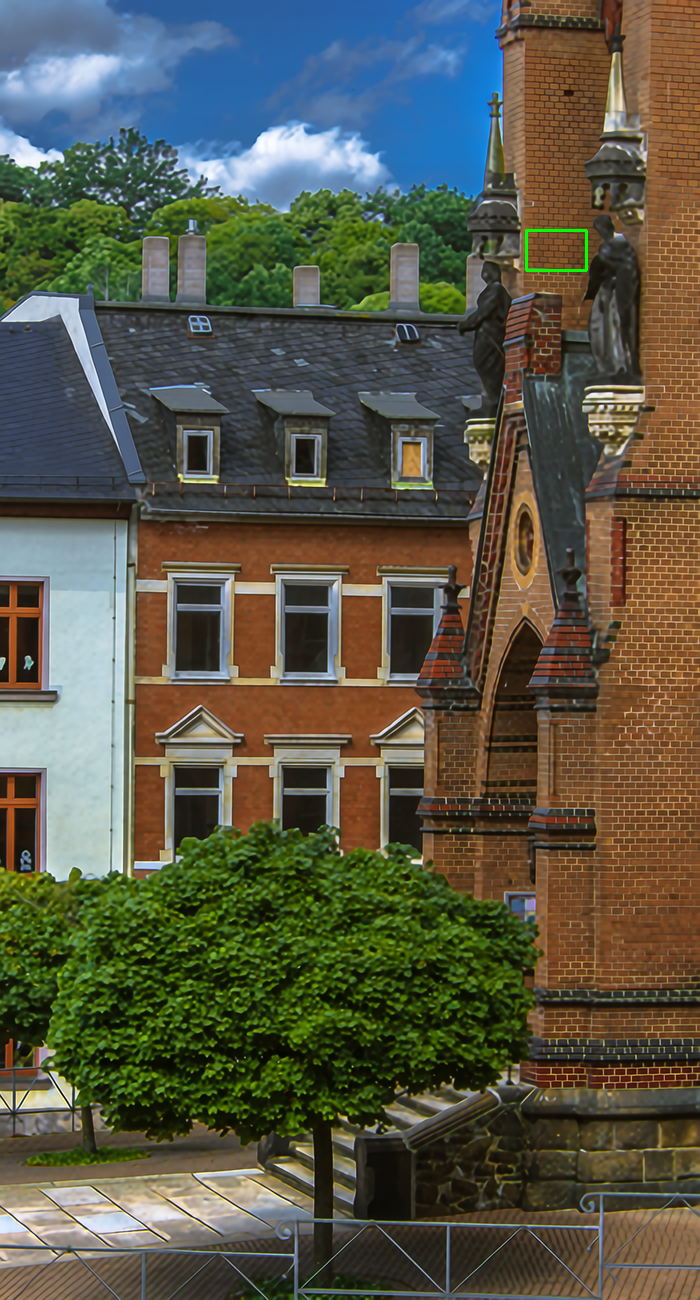}
\\
img\_0049
\end{tabular}
\end{adjustbox}
\hspace{-0.46cm}
\begin{adjustbox}{valign=t}
\begin{tabular}{cccccc}

\includegraphics[width=0.161\textwidth]{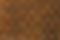} \hspace{-4mm} &
\includegraphics[width=0.161\textwidth]{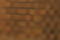} \hspace{-4mm} &
\includegraphics[width=0.161\textwidth]{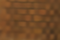} \hspace{-4mm} &
\includegraphics[width=0.161\textwidth]{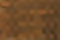} \hspace{-4mm} &
\includegraphics[width=0.161\textwidth]{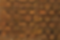} \hspace{-4mm}
\\

Bicubic \hspace{-4mm} &
RDN~\cite{zhang2018residual} \hspace{-4mm} &
RCAN~\cite{zhang2018image} \hspace{-4mm} &
StereoSR~\cite{jeon2018enhancing} \hspace{-4mm} &
SRRes+SAM~\cite{ying2020stereo} \hspace{-4mm} &
\\

\includegraphics[width=0.161\textwidth]{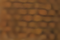} \hspace{-4mm} &
\includegraphics[width=0.161\textwidth]{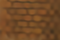} \hspace{-4mm} &
\includegraphics[width=0.161\textwidth]{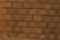} \hspace{-4mm} &
\includegraphics[width=0.161\textwidth]{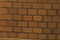} \hspace{-4mm}   &
\includegraphics[width=0.161\textwidth]{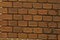} \hspace{-4mm}
\\

iPASSR~\cite{wang2021symmetric} \hspace{-4mm} &
SSRDE-FNet~\cite{dai2021feedback}  \hspace{-4mm} &
NAFSSR-L ~\cite{chu2022nafssr}  \hspace{-4mm} &
SCGLANet (\textbf{Ours})  \hspace{-4mm} &
Ground Truth \hspace{-4mm}
\\
\end{tabular}
\end{adjustbox}
\vspace{1mm}
\\
\end{tabular}
\caption{Visual results ($\times$4) achieved by different methods on the  Flickr1024~\cite{wang2019learning} dataset.
}
\label{fig:flickr1024}
\end{figure*} 

\begin{figure}
    \centering
    \includegraphics [width=0.48\textwidth]{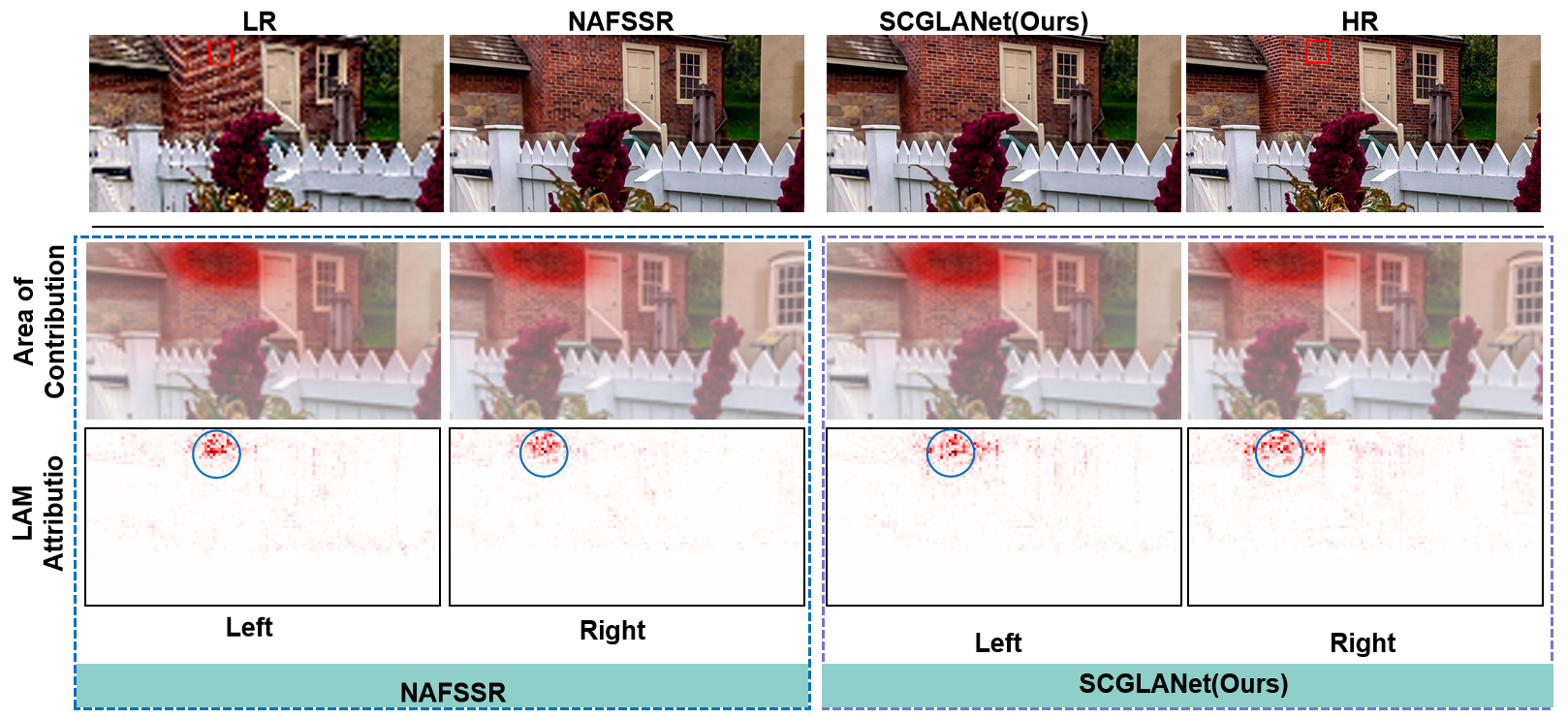}
    \caption{Comparison of the SR results and LAM attribution results of different SR networks. The LAM results visualize the importance of different pixel w.r.t. the SR results.}
    \label{fig:LAM}
\end{figure}

\subsection{Results}

\subsubsection{Classic Stereo Image Super-resolution Results}


\textbf{Quantitative Results.} This paper aims to thoroughly demonstrate the effectiveness of the proposed stereo image generator and expand upon the findings of our previous work \cite{Zhou2023Stereo}. To achieve this, a comprehensive comparison was performed on classic stereo super-resolution datasets. The compared methods include leading single-image super-resolution methods VDSR \cite{kim2016accurate}, EDSR \cite{lim2017enhanced}, RDN \cite{zhang2018residual}, and RCAN \cite{zhang2018image}, which were retrained by Wang et al. \cite{wang2021symmetric} using the Flickr1024 and Middlebury datasets. Additionally, several state-of-the-art stereo super-resolution algorithms, such as StereoSR \cite{jeon2018enhancing}, PASSRnet \cite{wang2019learning}, SRRes+SAM \cite{ying2020stereo}, IMSSRnet \cite{lei2020deep}, iPASSR \cite{wang2021symmetric}, SSRDE-FNet \cite{dai2021feedback}, and NAFSSR \cite{chu2022nafssr}, were included for comparison purposes. The results of the comparison are presented in \cref{tab:flickr1024_kitti_mid}.

As shown in \cref{tab:flickr1024_kitti_mid}, the term ``$Left$" refers to testing using only the left view, while ``$(Left+Right)/2$" represents the average results obtained from both the left and right views. To simplify the terminology, this paper defines these two testing modes as the Left View mode and the Average View mode, respectively. \cref{tab:flickr1024_kitti_mid} clearly demonstrates the effectiveness of the proposed method, illustrating its significant advantages over other competitive methods in x2 and x4 super-resolution. Specifically, in the Average View mode, the proposed method improves the PSNR metric by 0.45dB and 0.29dB compared to the state-of-the-art algorithm NAFSSR\_L on the in-domain datasets Middlebury and Flickr1024, respectively, for x2 super-resolution. Moreover, the proposed method achieves a PSNR improvement of 0.18dB for x4 super-resolution. Additionally, on the out-of-domain datasets KITTI2012 and KITTI2015, the proposed method achieves a PSNR increase of 0.08dB and 0.05dB, respectively. These results obviously demonstrate the efficacy of the proposed method. Furthermore, an intriguing observation is made in the out-of-domain dataset KITTI2015, where all previous state-of-the-art methods exhibit significant differences between the two testing modes. For instance, NAFSSR\_L achieves a PSNR of 31.25dB in the Average View mode, whereas it achieves only 30.46dB in the Left View mode, resulting in a difference of 0.79dB. However, the proposed method demonstrates a small discrepancy of only 0.09dB between the two view testing results, suggesting its robustness in handling inputs with varying disparities.


\cref{tab:pandf} provides a detailed comparison of parameters and FLOPs between the proposed method and the state-of-the-art method NAFSSR. In terms of x4 super-resolution, the parameters increase by approximately 17M, and the FLOPs increases by about 114G when comparing NAFSSR\_B to NAFSSR\_L. However, there is only a marginal gain of 0.1dB and 0.05dB in PSNR on the in-domain dataset Flickr1024 and the out-of-domain dataset KITTI2012, respectively. Conversely, when comparing NAFSSR\_L to the proposed method, the parameters increase by 1.46M, and the FLOPs increases by 23.8G. Remarkably, this results in a PSNR improvement of 0.29dB and 0.11dB on Flickr1024 and KITTI2012, respectively, indicating the efficiency of the proposed method.

\begin{table}
\centering
\caption{Comparison results of parameters and FLOPs of different methods.}
\label{tab:pandf}
\resizebox{0.35\textwidth}{!}{
\begin{tabular}{llll}
\hline
Model & Scale & \#P(M) & GFLOPs \\ \hline
NAFSSR\_B & X2 & 6.77 & 45.1 \\
NAFSSR\_L & X2 & 23.79 & 158.7 \\
SCGLANet & X2 & 25.25 & 182.5 \\ \hline
NAFSSR\_B & X4 & 6.80 & 45.3 \\
NAFSSR\_L & X4 & 23.83 & 159.0 \\
SCGLANet & X4 & 25.29 & 182.8 \\ \hline
\end{tabular}}
\end{table}



\textbf{Qualitative Results} \cref{fig:flickr1024} presents the qualitative comparison results of the proposed method and other methods on the Flickr1024 dataset. It is evident that the proposed method exhibits high restoration capability for severely damaged images, especially those with strong textures. The restored textures appear more natural and realistic in comparison to the other methods. This is particularly noteworthy for images with repetitive textures, such as the ``brick wall" image. Despite the input image's texture being indistinguishable, the proposed method is able to recover reasonable textures. More qualitative results can be found in supplementary material. 
To further investigate the reasons behind the effectiveness of the proposed method, this paper utilized the interpretability tool known as local attribution map (LAM) \cite{gu2021interpreting}. This map was utilized for the attribution understanding of two approaches, namely NAFSSR and SCGLANet. The attribution results can be seen in \cref{fig:LAM}. Upon observing the vicinity of the blue circle in the figure, it becomes evident that the proposed method possesses a wider focus range compared to NAFSSR and exhibits a more precise focus area.

\begin{figure*}[htb]
\scriptsize
\centering
\begin{tabular}{cc}
\hspace{-0.4cm}
\begin{adjustbox}{valign=t}
\begin{tabular}{c}
\includegraphics[width=0.12\textwidth,height=0.13\textheight]{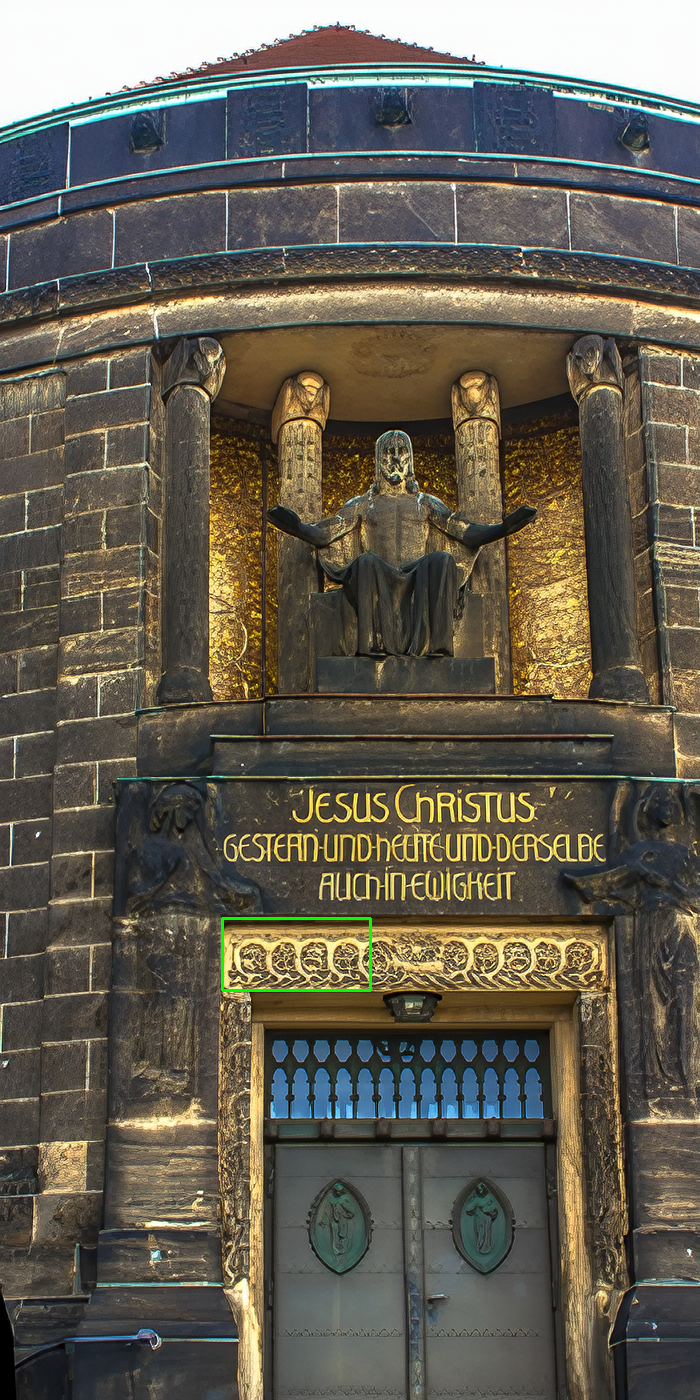}
\\
img\_0015
\end{tabular}
\end{adjustbox}
\hspace{-0.46cm}
\begin{adjustbox}{valign=t}
\begin{tabular}{cccccc}
\includegraphics[width=0.161\textwidth]{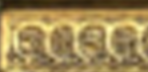} \hspace{-4mm} &
\includegraphics[width=0.161\textwidth]{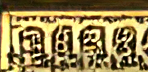} \hspace{-4mm} &
\includegraphics[width=0.161\textwidth]{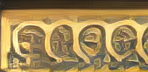} \hspace{-4mm} &
\includegraphics[width=0.161\textwidth]{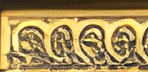} \hspace{-4mm} &
\includegraphics[width=0.161\textwidth]{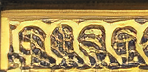} \hspace{-4mm}
\\

Bicubic \hspace{-4mm} &
RealSR(CVPRW20')~\cite{ji2020real} \hspace{-4mm} &
BSRGAN(ICCV21')~\cite{zhang2021designing} \hspace{-4mm} &
SwinIR(ICCVW21')~\cite{liang2021swinir} \hspace{-4mm} &
RealESRGAN(CVPR21')~\cite{wang2021real} \hspace{-4mm} &
\\

\includegraphics[width=0.161\textwidth]{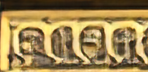} \hspace{-4mm} &
\includegraphics[width=0.161\textwidth]{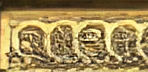} \hspace{-4mm} &
\includegraphics[width=0.161\textwidth]{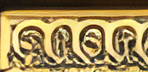} \hspace{-4mm} &
\includegraphics[width=0.161\textwidth]{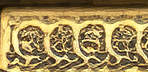} \hspace{-4mm}   &
\includegraphics[width=0.161\textwidth]{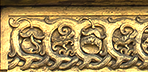} \hspace{-4mm}
\\

PDM-SRGAN(CVPR22')~\cite{luo2022learning} \hspace{-4mm} &
DASR(ECCV22')~\cite{liang2022efficient}  \hspace{-4mm} &
MMRealSR(ECCV22')~\cite{mou2022metric}  \hspace{-4mm} &
RealSCGLAGAN(\textbf{Ours})    \hspace{-4mm} &
Ground Truth \hspace{-4mm}
\\
\end{tabular}
\end{adjustbox}
\vspace{1mm}
\\

\hspace{-0.4cm}
\begin{adjustbox}{valign=t}
\begin{tabular}{c}
\includegraphics[width=0.12\textwidth,height=0.173\textheight]{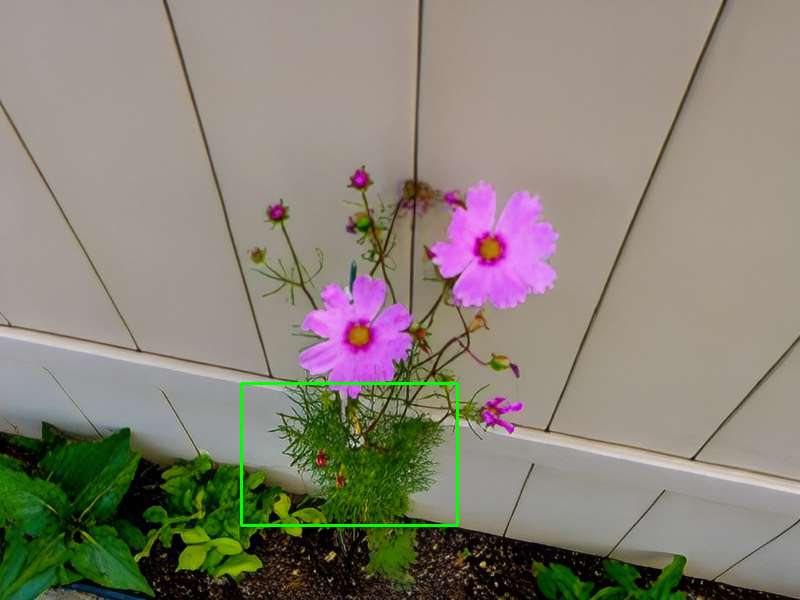}
\\
img\_0090
\end{tabular}
\end{adjustbox}
\hspace{-0.46cm}
\begin{adjustbox}{valign=t}
\begin{tabular}{cccccc}

\includegraphics[width=0.161\textwidth]{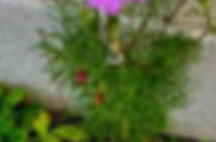} \hspace{-4mm} &
\includegraphics[width=0.161\textwidth]{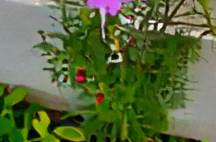} \hspace{-4mm} &
\includegraphics[width=0.161\textwidth]{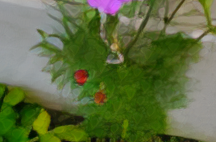} \hspace{-4mm} &
\includegraphics[width=0.161\textwidth]{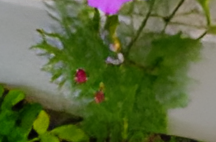} \hspace{-4mm} &
\includegraphics[width=0.161\textwidth]{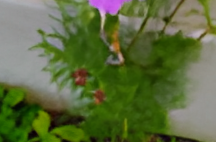} \hspace{-4mm}
\\

Bicubic \hspace{-4mm} &
RealSR(CVPRW20')~\cite{ji2020real} \hspace{-4mm} &
BSRGAN(ICCV21')~\cite{zhang2021designing} \hspace{-4mm} &
SwinIR(ICCVW21')~\cite{liang2021swinir} \hspace{-4mm} &
RealESRGAN(CVPR21')~\cite{wang2021real} \hspace{-4mm} &
\\

\includegraphics[width=0.161\textwidth]{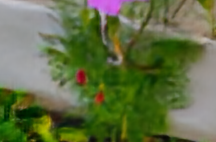} \hspace{-4mm} &
\includegraphics[width=0.161\textwidth]{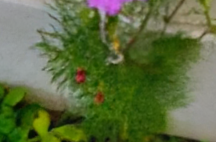} \hspace{-4mm} &
\includegraphics[width=0.161\textwidth]{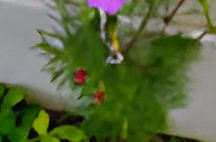} \hspace{-4mm} &
\includegraphics[width=0.161\textwidth]{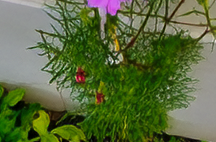} \hspace{-4mm}   &
\includegraphics[width=0.161\textwidth]{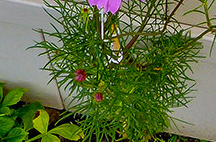} \hspace{-4mm}
\\

PDM-SRGAN(CVPR22')~\cite{luo2022learning} \hspace{-4mm} &
DASR(ECCV22')~\cite{liang2022efficient}  \hspace{-4mm} &
MMRealSR(ECCV22')~\cite{mou2022metric}  \hspace{-4mm} &
RealSCGLAGAN(\textbf{Ours})    \hspace{-4mm} &
Ground Truth \hspace{-4mm}
\\
\end{tabular}
\end{adjustbox}
\vspace{1mm}
\\
\end{tabular}
\caption{Visual results ($\times$4) achieved by different methods on the  Flickr1024RS~\cite{wang2019learning} dataset.
}
\label{fig:flickr1024RS}
\end{figure*}

\begin{figure*}[htb]
\scriptsize
\centering
\begin{tabular}{cc}
\hspace{-0.4cm}
\begin{adjustbox}{valign=t}
\begin{tabular}{c}
\includegraphics[width=0.12\textwidth,height=0.0743\textheight]{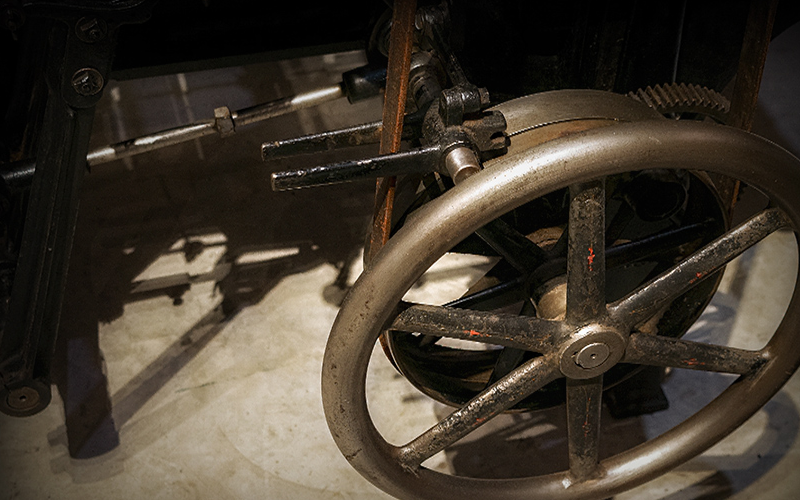}
\\
Left
\\
\includegraphics[width=0.12\textwidth,height=0.0743\textheight]{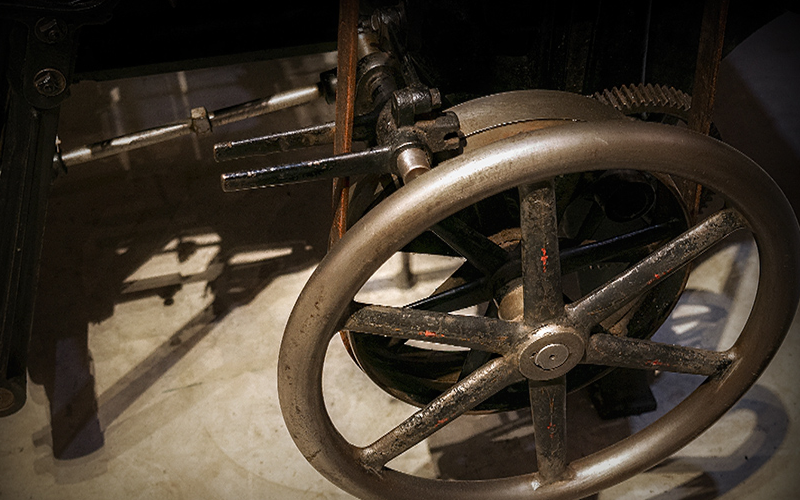}
\\
Right
\end{tabular}
\end{adjustbox}
\hspace{-0.46cm}
\begin{adjustbox}{valign=t}
\begin{tabular}{cccccc}
\includegraphics[width=0.161\textwidth]{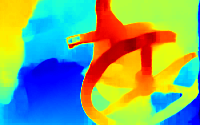} \hspace{-4mm} &
\includegraphics[width=0.161\textwidth]{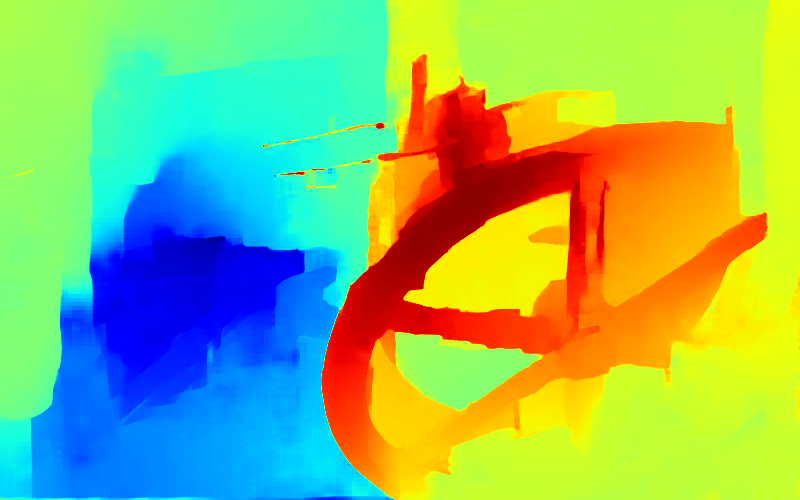} \hspace{-4mm} &
\includegraphics[width=0.161\textwidth]{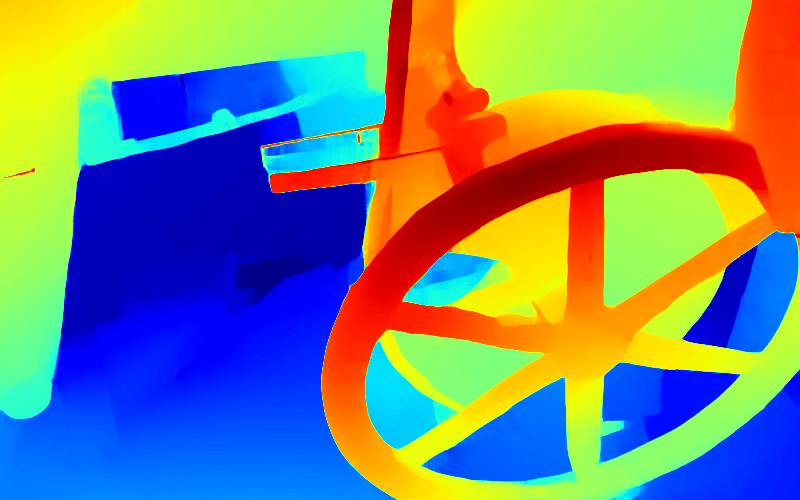} \hspace{-4mm} &
\includegraphics[width=0.161\textwidth]{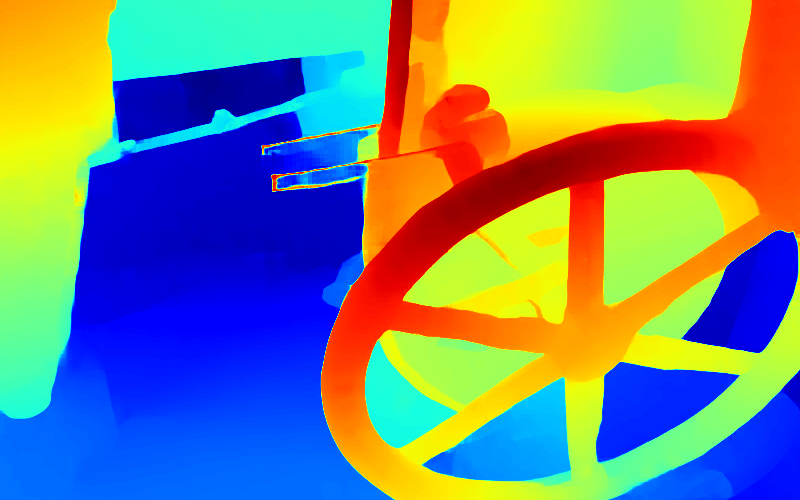} \hspace{-4mm} &
\includegraphics[width=0.161\textwidth]{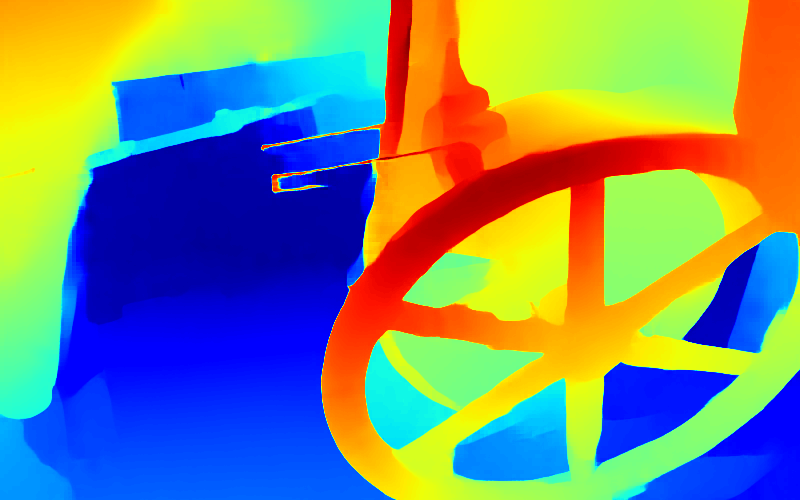} \hspace{-4mm}
\\

Bicubic \hspace{-4mm} &
RealSR(CVPRW20')~\cite{ji2020real} \hspace{-4mm} &
BSRGAN(ICCV21')~\cite{zhang2021designing} \hspace{-4mm} &
SwinIR(ICCVW21')~\cite{liang2021swinir} \hspace{-4mm} &
RealESRGAN(CVPR21')~\cite{wang2021real} \hspace{-4mm} &
\\

\includegraphics[width=0.161\textwidth]{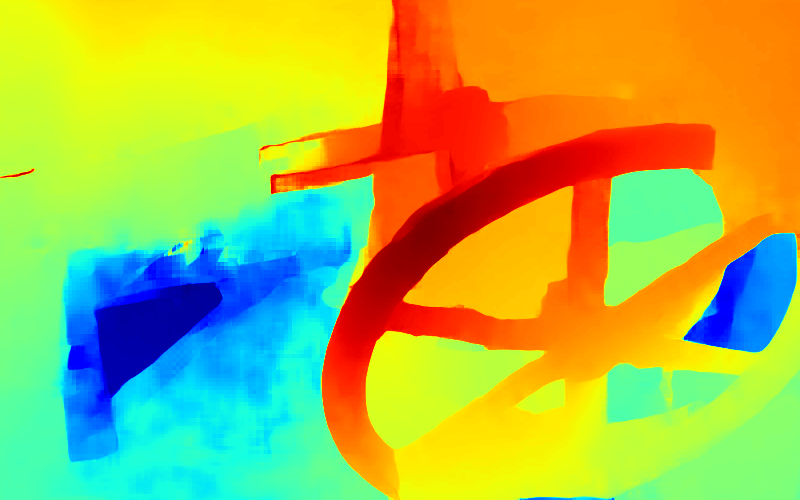} \hspace{-4mm} &
\includegraphics[width=0.161\textwidth]{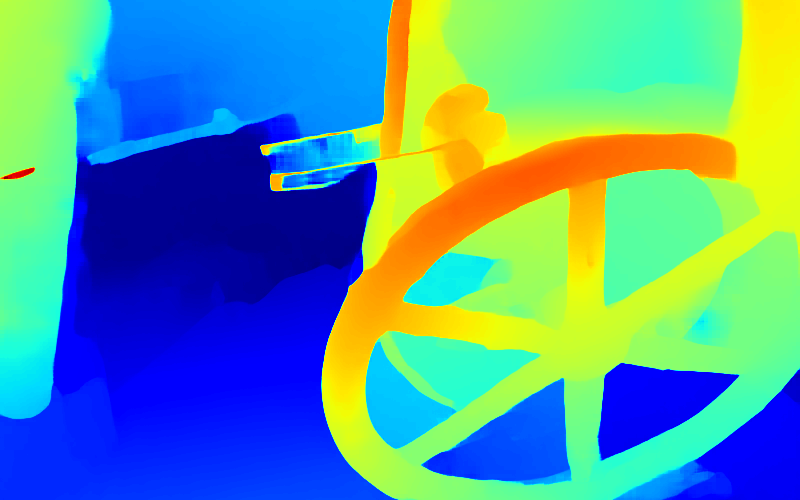} \hspace{-4mm} &
\includegraphics[width=0.161\textwidth]{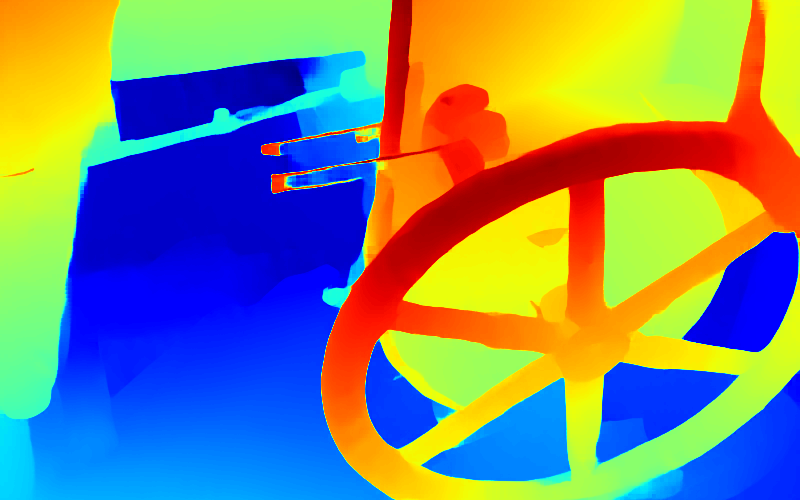} \hspace{-4mm} &
\includegraphics[width=0.161\textwidth]{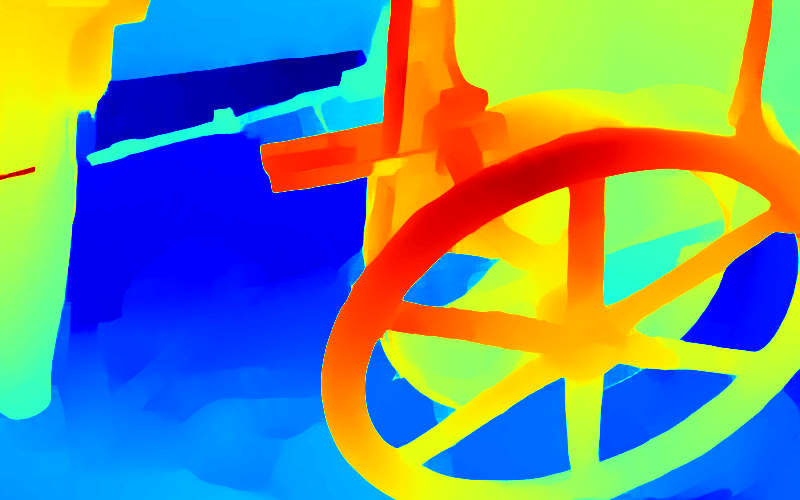} \hspace{-4mm} &
\includegraphics[width=0.161\textwidth]{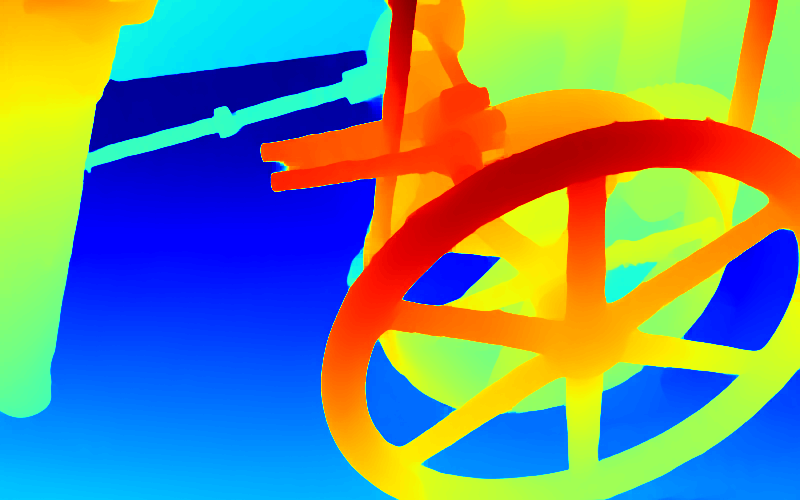} \hspace{-4mm}
\\

PDM-SRGAN(CVPR22')~\cite{luo2022learning} \hspace{-4mm} &
DASR(ECCV22')~\cite{liang2022efficient}  \hspace{-4mm} &
MMRealSR(ECCV22')~\cite{mou2022metric}  \hspace{-4mm} &
RealSCGLAGAN(\textbf{Ours})    \hspace{-4mm} &
Ground Truth \hspace{-4mm}
\\
\end{tabular}
\end{adjustbox}
\vspace{1mm}
\\

\hspace{-0.4cm}
\begin{adjustbox}{valign=t}
\begin{tabular}{c}
\includegraphics[width=0.12\textwidth,height=0.148\textheight]{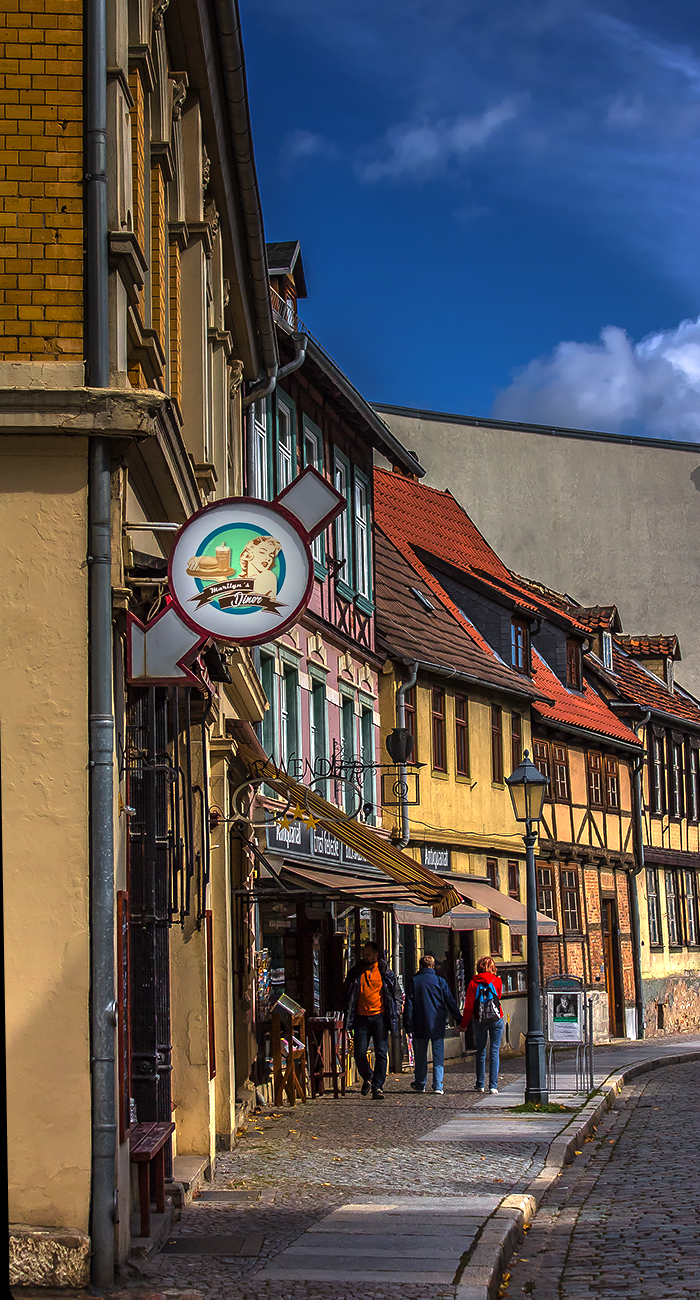}
\\
Left
\\
\includegraphics[width=0.12\textwidth,height=0.148\textheight]{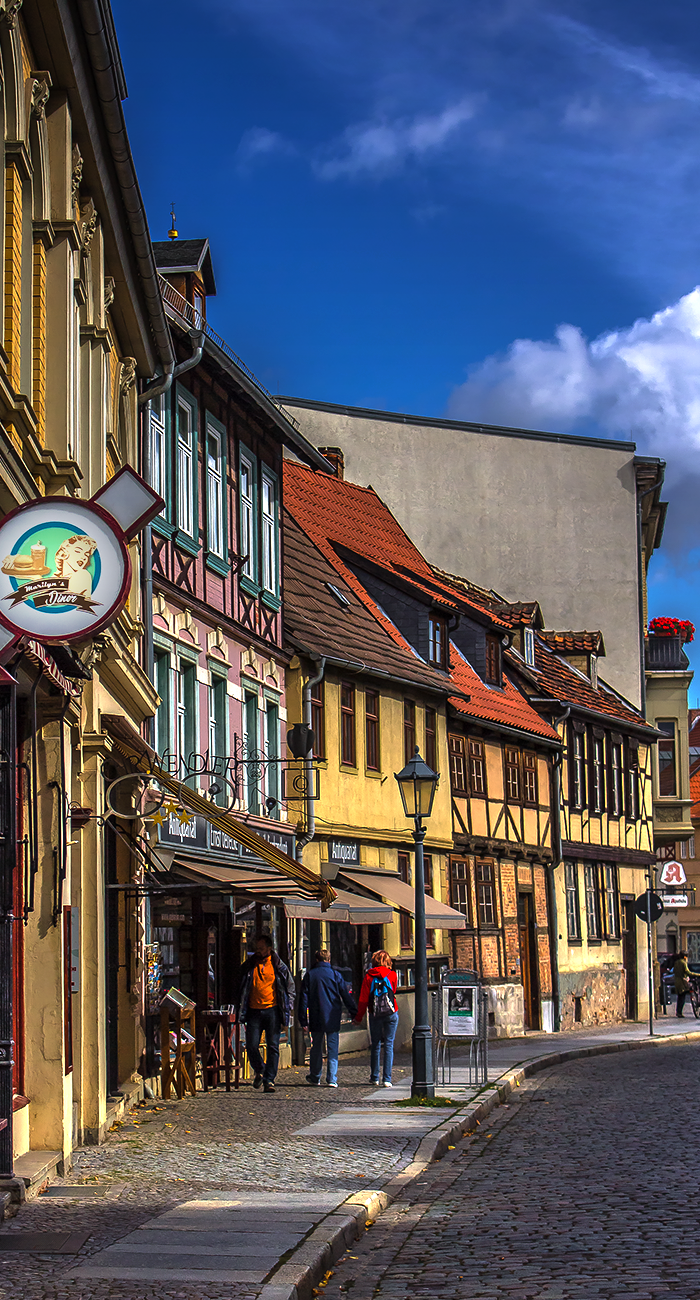}
\\
Right
\end{tabular}
\end{adjustbox}
\hspace{-0.46cm}
\begin{adjustbox}{valign=t}
\begin{tabular}{cccccc}

\includegraphics[width=0.161\textwidth,height=0.2\textwidth]{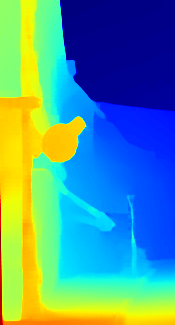} \hspace{-4mm} &
\includegraphics[width=0.161\textwidth,height=0.2\textwidth]{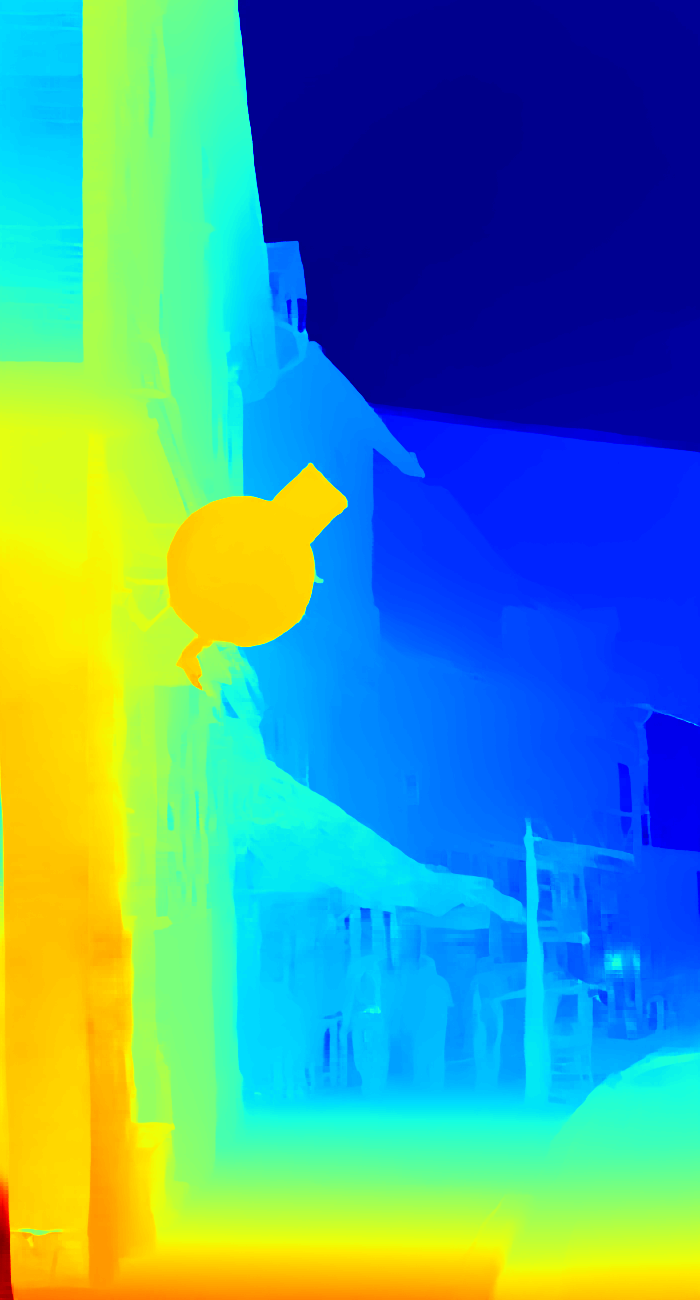} \hspace{-4mm} &
\includegraphics[width=0.161\textwidth,height=0.2\textwidth]{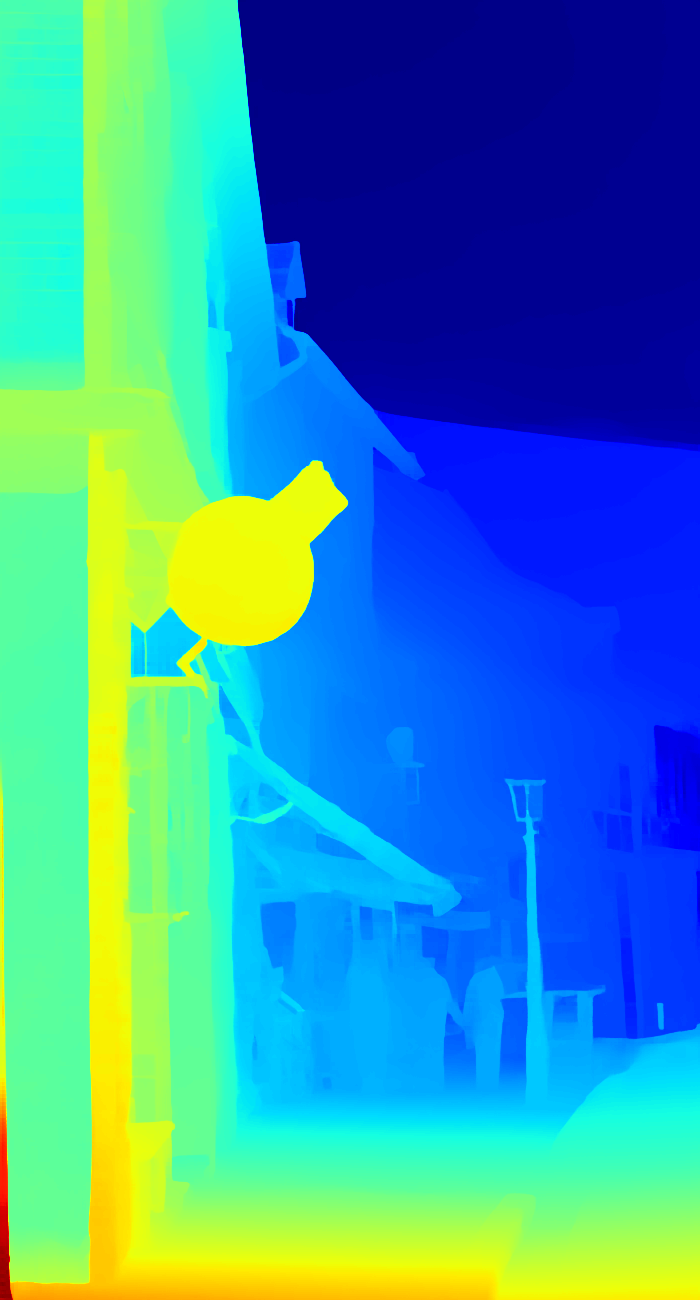} \hspace{-4mm} &
\includegraphics[width=0.161\textwidth,height=0.2\textwidth]{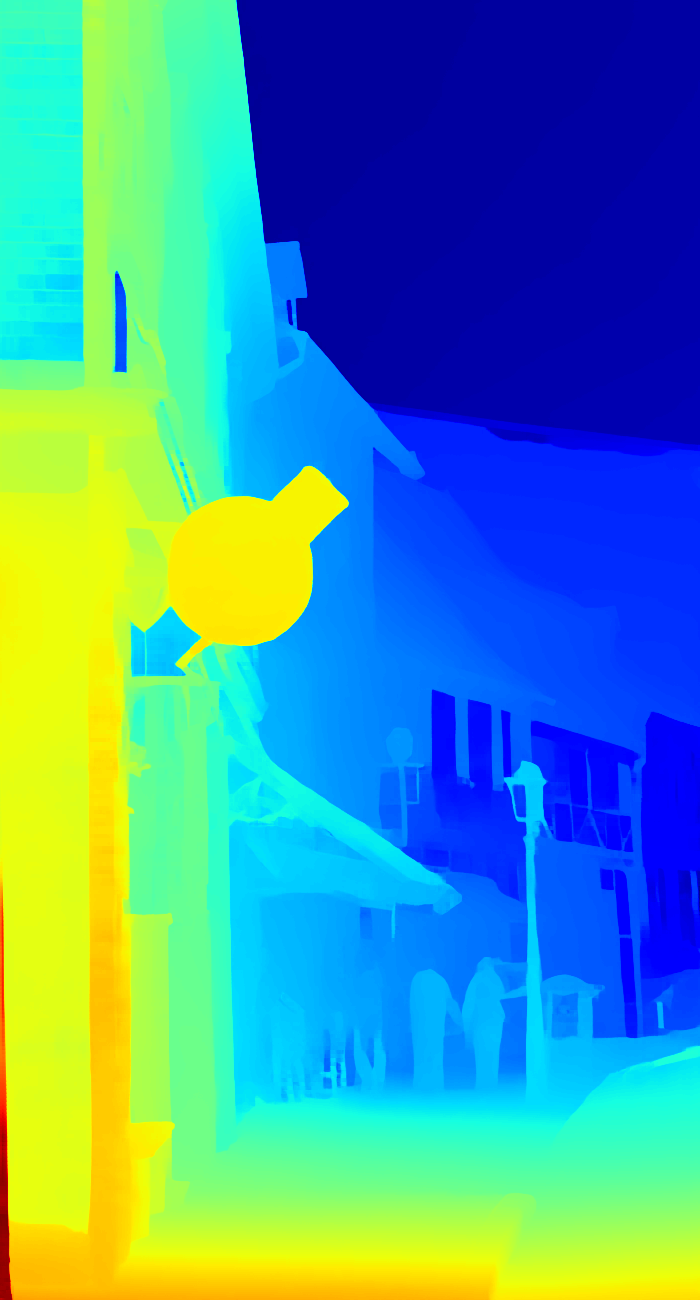} \hspace{-4mm} &
\includegraphics[width=0.161\textwidth,height=0.2\textwidth]{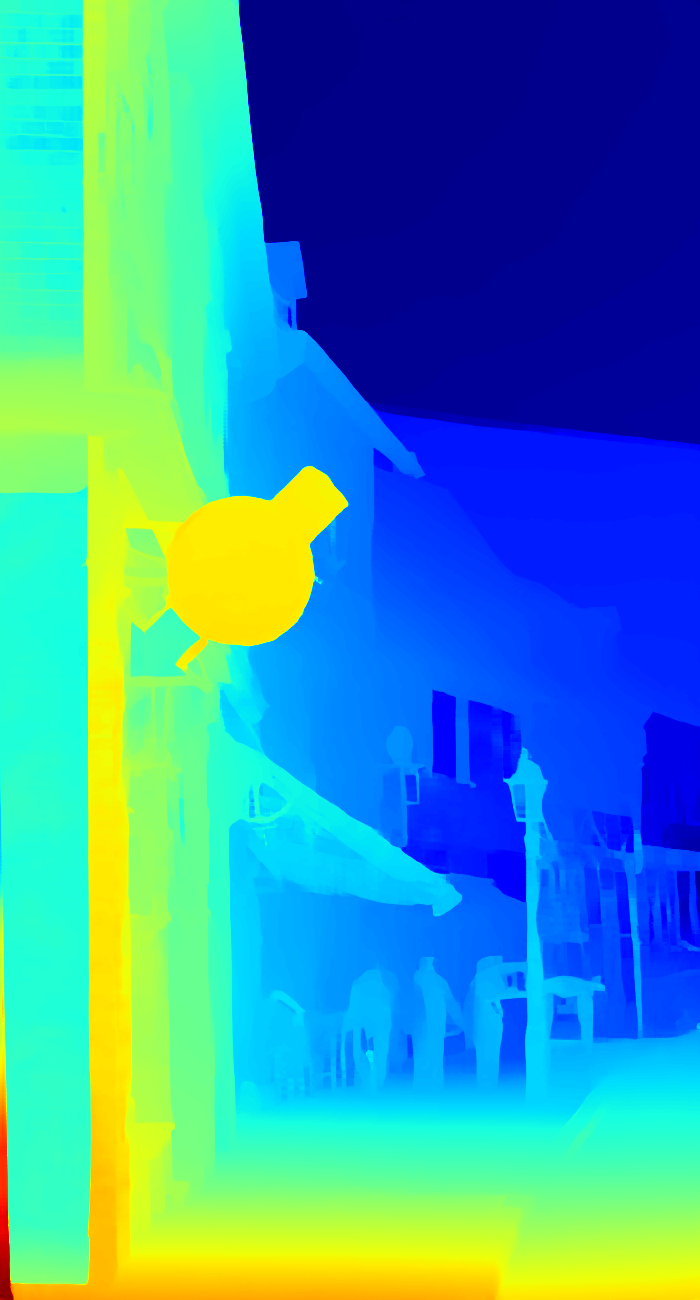} \hspace{-4mm}
\\

Bicubic \hspace{-4mm} &
RealSR(CVPRW20')~\cite{ji2020real} \hspace{-4mm} &
BSRGAN(ICCV21')~\cite{zhang2021designing} \hspace{-4mm} &
SwinIR(ICCVW21')~\cite{liang2021swinir} \hspace{-4mm} &
RealESRGAN(CVPR21')~\cite{wang2021real} \hspace{-4mm} &
\\

\includegraphics[width=0.161\textwidth,height=0.2\textwidth]{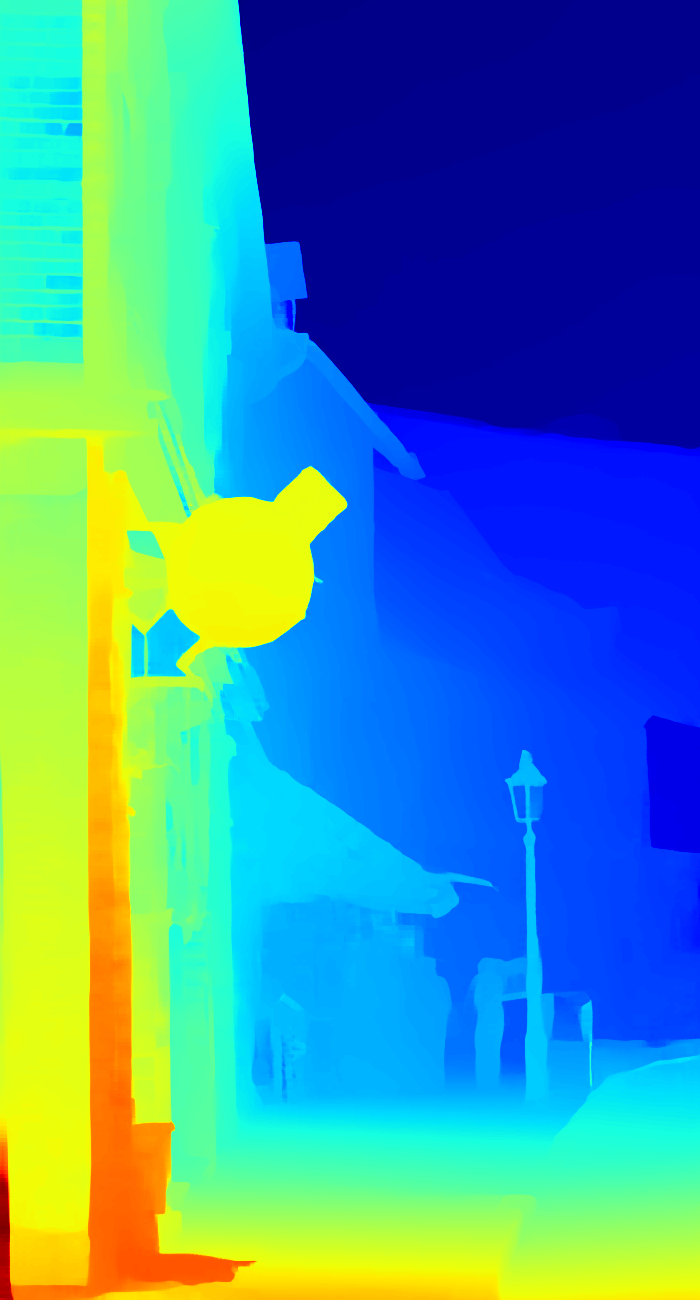} \hspace{-4mm} &
\includegraphics[width=0.161\textwidth,height=0.2\textwidth]{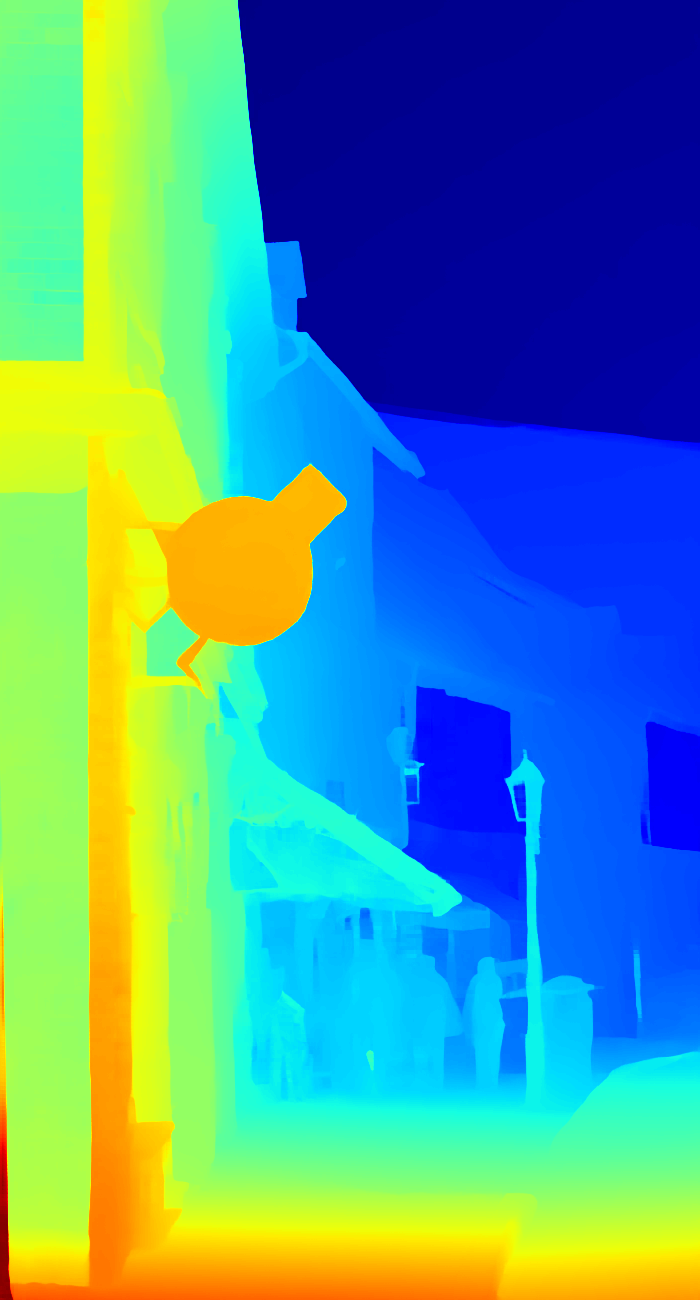} \hspace{-4mm} &
\includegraphics[width=0.161\textwidth,height=0.2\textwidth]{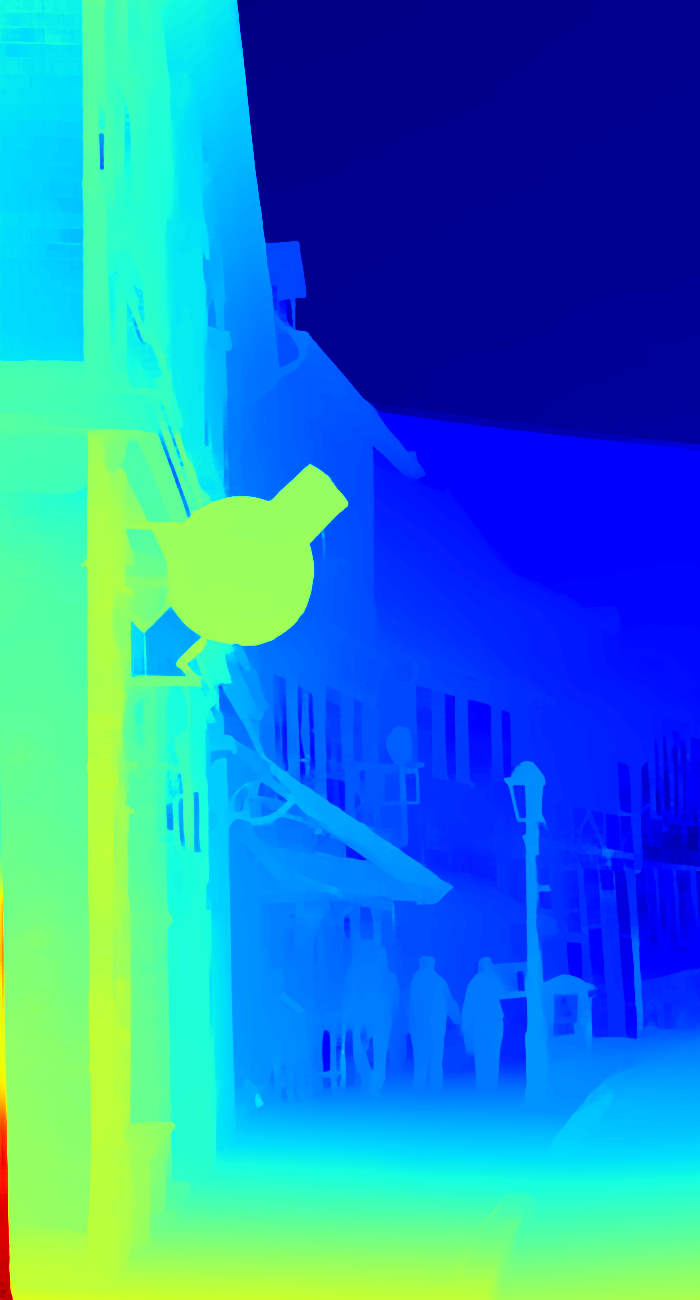} \hspace{-4mm} &
\includegraphics[width=0.161\textwidth,height=0.2\textwidth]{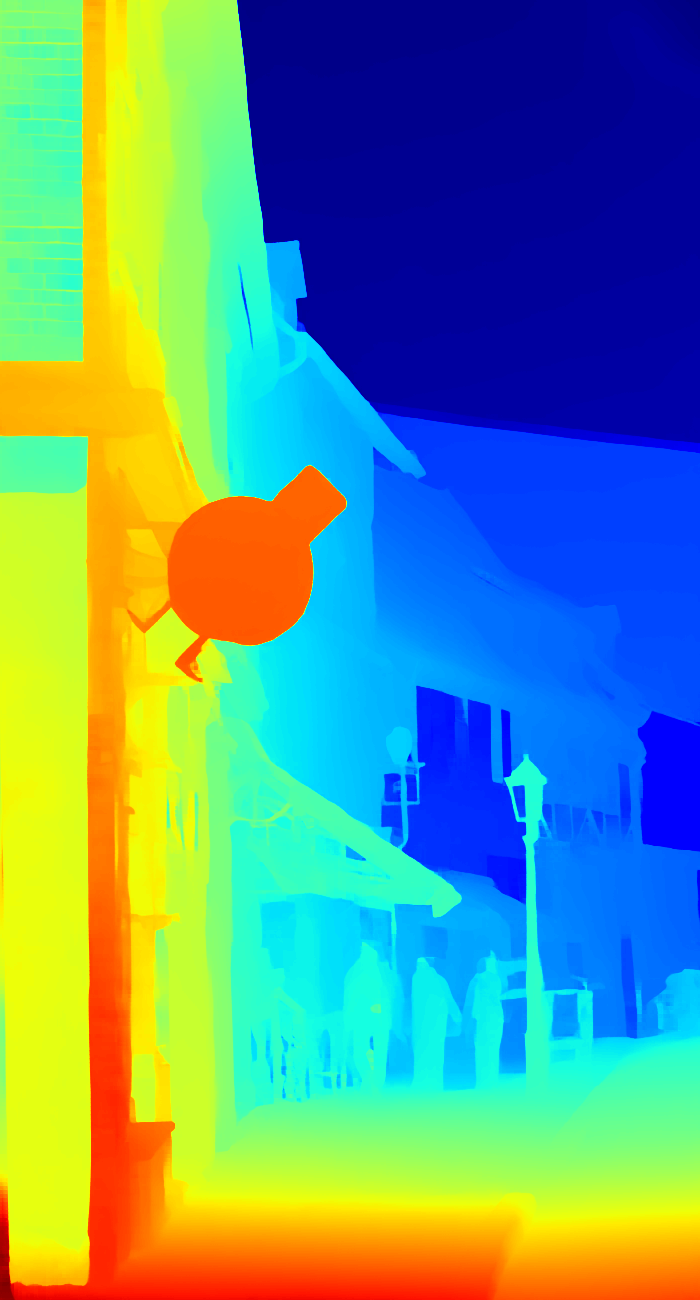} \hspace{-4mm} &
\includegraphics[width=0.161\textwidth,height=0.2\textwidth]{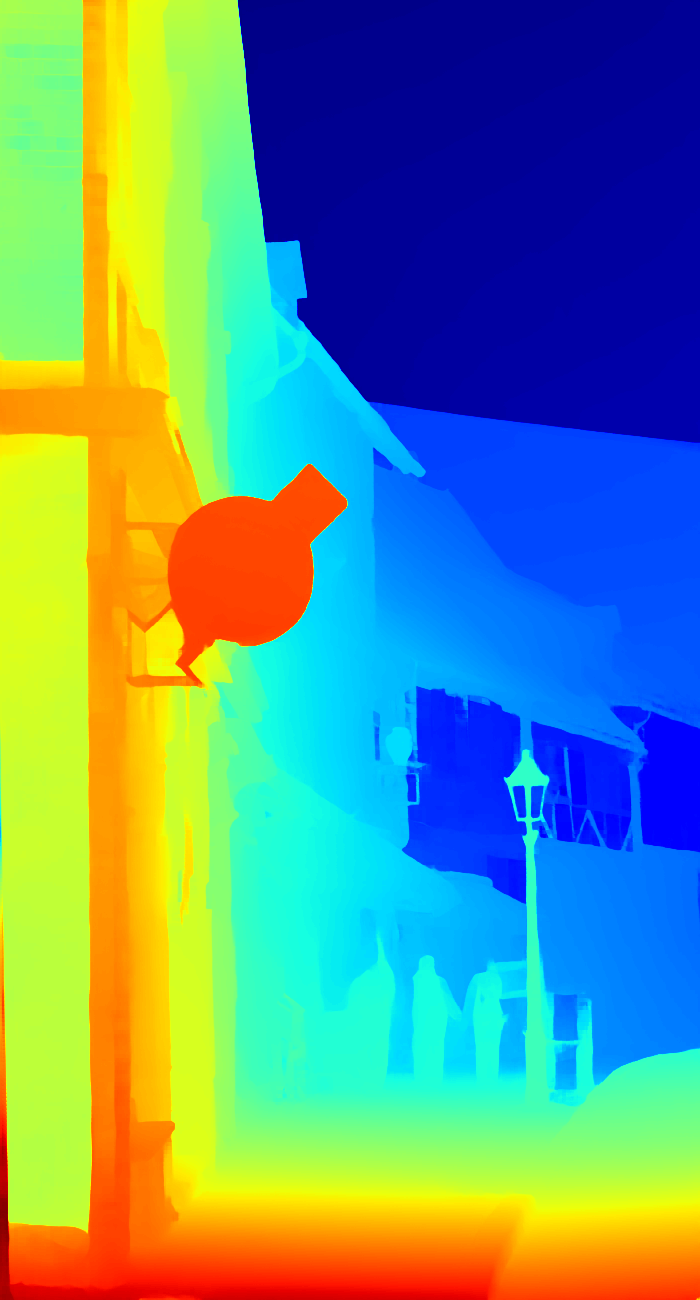} \hspace{-4mm}
\\

PDM-SRGAN(CVPR22')~\cite{luo2022learning} \hspace{-4mm} &
DASR(ECCV22')~\cite{liang2022efficient}  \hspace{-4mm} &
MMRealSR(ECCV22')~\cite{mou2022metric}  \hspace{-4mm} &
RealSCGLAGAN(\textbf{Ours})   \hspace{-4mm} &
Ground Truth \hspace{-4mm}
\\
\end{tabular}
\end{adjustbox}
\vspace{1mm}
\\
\end{tabular}
\caption{The visual results of disparity estimated images achieved by different methods on the Flickr1024RS~\cite{wang2019learning} dataset.
}
\label{fig:flickr1024RSDis}
\end{figure*} 

\begin{table}
\centering
\caption{Quantitative results achieved by different methods on the Flickr1024RS datasets.
} \label{tab:filckr1024rs}
\resizebox{0.48\textwidth}{!}{
\begin{tabular}{lllll}
\hline
\multicolumn{1}{c}{\multirow{2}{*}{Methos}} & \multicolumn{4}{c}{$\text{Results}^{*}$} \\ \cline{2-5}
\multicolumn{1}{c}{} & PSNR (RGB)$\uparrow$ & SSIM $\uparrow$ & LPIPS $\downarrow$ & MADE $\downarrow$ \\ \hline
RealSR(CVPRW20')~\cite{ji2020real} & 18.20 & 0.4726 & 0.4048 & 6.93 \\
BSRGAN(ICCV21')~\cite{zhang2021designing} & 20.82 & 0.5694 & 0.3233 & 3.49 \\
SwinIR(ICCVW21')~\cite{liang2021swinir} & 20.43 & 0.5678 & 0.3016 & 3.54 \\
RealESRGAN(CVPR21')~\cite{wang2021real} & 20.50 & 0.5713 & 0.3001 & 3.65 \\
PDM-SRGAN(CVPR22')~\cite{luo2022learning} & 19.06 & 0.5085 & 0.3891 & 4.14 \\
DASR(ECCV22')~\cite{liang2022efficient} & \textbf{21.23} & \underline{0.5981} & \underline{0.2878} & \underline{3.01} \\
MMRealSR(ECCV22')~\cite{mou2022metric} & 20.00 & 0.5792 & 0.2956 & 3.41 \\
RealSCGLAGAN(Ours) & \underline{21.06} & \textbf{0.6235} & \textbf{0.2098} & \textbf{2.20} \\ \hline
\multicolumn{5}{l}{$*$  The best results are in \textbf{bold faces} and the second results are in \underline{underline}.} \\
\end{tabular}}
\end{table}

\begin{figure*}[t]
\scriptsize
\centering
\begin{tabular}{cc}

\hspace{-0.4cm}
\begin{adjustbox}{valign=t}
\begin{tabular}{c}
\includegraphics[width=0.12\textwidth,height=0.115\textheight]{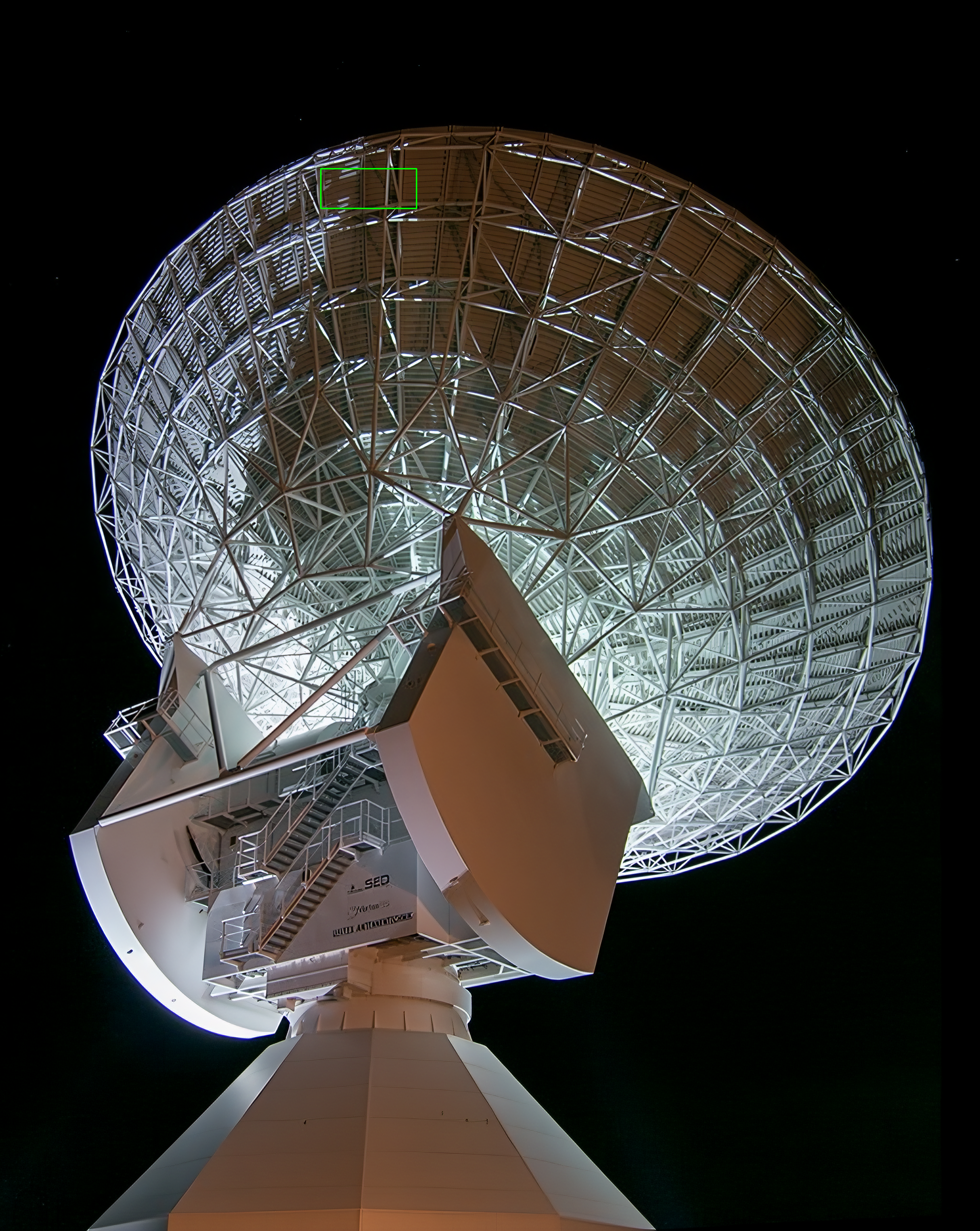}
\\
\end{tabular}
\end{adjustbox}
\hspace{-0.46cm}
\begin{adjustbox}{valign=t}
\begin{tabular}{cccccc}

\includegraphics[width=0.161\textwidth]{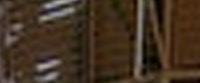} \hspace{-4mm} &
\includegraphics[width=0.161\textwidth]{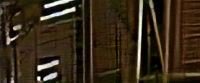} \hspace{-4mm} &
\includegraphics[width=0.161\textwidth]{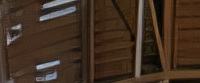} \hspace{-4mm} &
\includegraphics[width=0.161\textwidth]{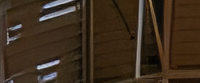} \hspace{-4mm} &
\includegraphics[width=0.161\textwidth]{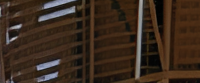} \hspace{-4mm}
\\

Bicubic \hspace{-4mm} &
RealSR(CVPRW20')~\cite{ji2020real} \hspace{-4mm} &
BSRGAN(ICCV21')~\cite{zhang2021designing} \hspace{-4mm} &
SwinIR(ICCVW21')~\cite{liang2021swinir} \hspace{-4mm} &
RealESRGAN(CVPR21')~\cite{wang2021real} \hspace{-4mm} &
\\

\includegraphics[width=0.161\textwidth]{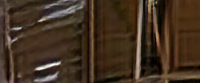} \hspace{-4mm} &
\includegraphics[width=0.161\textwidth]{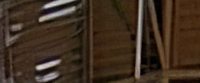} \hspace{-4mm} &
\includegraphics[width=0.161\textwidth]{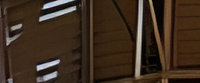} \hspace{-4mm} &
\includegraphics[width=0.161\textwidth]{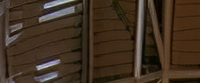} \hspace{-4mm} &
\includegraphics[width=0.161\textwidth]{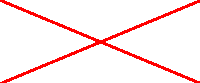} \hspace{-4mm}
\\

PDM-SRGAN(CVPR22')~\cite{luo2022learning} \hspace{-4mm} &
DASR(ECCV22')~\cite{liang2022efficient}  \hspace{-4mm} &
MMRealSR(ECCV22')~\cite{mou2022metric}  \hspace{-4mm} &
RealSCGLAGAN(\textbf{Ours})     \hspace{-4mm} &
Ground Truth \hspace{-4mm}
\\
\end{tabular}
\end{adjustbox}
\vspace{1mm}
\\

\hspace{-0.4cm}
\begin{adjustbox}{valign=t}
\begin{tabular}{c}
\includegraphics[width=0.12\textwidth,height=0.165\textheight]{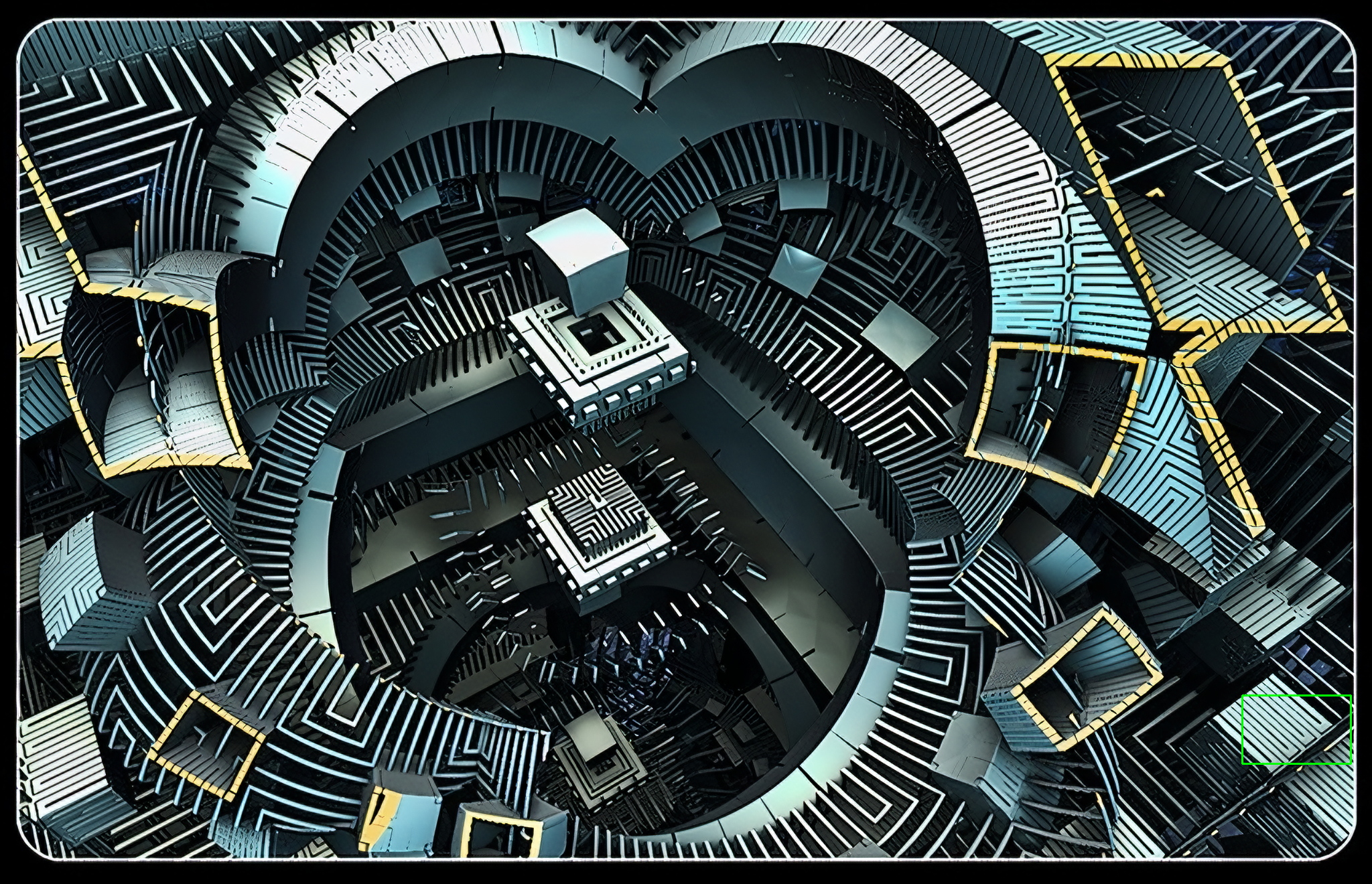}
\\
\end{tabular}
\end{adjustbox}
\hspace{-0.46cm}
\begin{adjustbox}{valign=t}
\begin{tabular}{cccccc}

\includegraphics[width=0.161\textwidth]{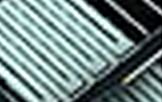} \hspace{-4mm} &
\includegraphics[width=0.161\textwidth]{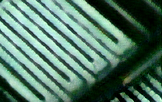} \hspace{-4mm} &
\includegraphics[width=0.161\textwidth]{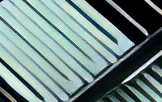} \hspace{-4mm} &
\includegraphics[width=0.161\textwidth]{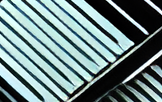} \hspace{-4mm} &
\includegraphics[width=0.161\textwidth]{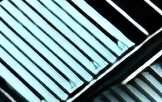} \hspace{-4mm}
\\

Bicubic \hspace{-4mm} &
RealSR(CVPRW20')~\cite{ji2020real} \hspace{-4mm} &
BSRGAN(ICCV21')~\cite{zhang2021designing} \hspace{-4mm} &
SwinIR(ICCVW21')~\cite{liang2021swinir} \hspace{-4mm} &
RealESRGAN(CVPR21')~\cite{wang2021real} \hspace{-4mm} &
\\

\includegraphics[width=0.161\textwidth]{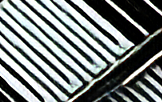} \hspace{-4mm} &
\includegraphics[width=0.161\textwidth]{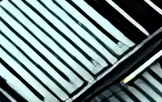} \hspace{-4mm} &
\includegraphics[width=0.161\textwidth]{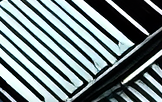} \hspace{-4mm} &
\includegraphics[width=0.161\textwidth]{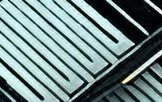} \hspace{-4mm} &
\includegraphics[width=0.161\textwidth]{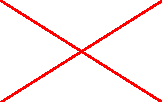} \hspace{-4mm}
\\

PDM-SRGAN(CVPR22')~\cite{luo2022learning} \hspace{-4mm} &
DASR(ECCV22')~\cite{liang2022efficient}  \hspace{-4mm} &
MMRealSR(ECCV22')~\cite{mou2022metric}  \hspace{-4mm} &
RealSCGLAGAN(\textbf{Ours})    \hspace{-4mm} &
Ground Truth \hspace{-4mm}
\\
\end{tabular}
\end{adjustbox}
\vspace{1mm}
\end{tabular}
\caption{Visual results ($\times$4) achieved by different methods on the StereoWeb20 dataset.}
\label{fig:StereoWeb20}
\end{figure*}

\begin{table}
\centering
\caption{Quantitative results achieved by different methods on the StereoWeb20 datasets.
} \label{tab:stereoweb20}
\resizebox{0.35\textwidth}{!}{
\begin{tabular}{lll}
\hline
\multicolumn{1}{c}{\multirow{2}{*}{Methods}} & \multicolumn{2}{c}{$\text{Results}^{*}$ } \\ \cline{2-3}
\multicolumn{1}{c}{} & NRQM $\uparrow$ & PI $\downarrow$  \\ \hline
RealSR(CVPRW20')~\cite{ji2020real} & 6.3145 & \textbf{3.9070} \\
BSRGAN(ICCV21')~\cite{zhang2021designing} & 6.4409 & 3.9703  \\
SwinIR(ICCVW21')~\cite{liang2021swinir} & 6.3270& 4.0468  \\
RealESRGAN(CVPR21')~\cite{wang2021real} & 6.3258 & 4.0006  \\
PDM-SRGAN(CVPR22')~\cite{luo2022learning} & 5.5687 & 4.7678  \\
DASR(ECCV22')~\cite{liang2022efficient} & 5.9030 & 4.1032 \\
MMRealSR(ECCV22')~\cite{mou2022metric} & \underline{6.3569} & 4.0107\\
RealSCGLAGAN(Ours) & \textbf{6.5915} & \underline{3.9090} \\ \hline
\multicolumn{3}{l}{$*$ The best results are in \textbf{bold faces} and the second }\\
\multicolumn{3}{l}{\hspace{1.8mm} results are in \underline{underline}.}
\end{tabular}}
\end{table}

\subsubsection{Real-World Stereo Super-Resolution Performance Comparison}

\textbf{Flickr1024RS Dataset.} To the best of our knowledge, there are currently no existing real-world stereo-image super-resolution methods. Therefore, in this study, we selected competitive single-image super-resolution methods for comparison, including RealSR \cite{ji2020real}, BSRGAN \cite{zhang2021designing}, RealESRGAN \cite{wang2021real}, SwinIR \cite{liang2021swinir}, PDM-SRGAN \cite{luo2022learning}, DASR \cite{liang2022efficient}, and MMRealSR \cite{mou2022metric}. The final quantitative results are presented in \cref{tab:filckr1024rs}. It is evident from the table that RealSR exhibits significant differences in performance compared to other real-world single-image super-resolution methods, particularly in terms of high MADE, which indicates severe disparity errors in the super-resolved results produced by RealSR. This is mainly attributed to the insufficient robustness of its degradation model and the lack of stereo information, resulting in a substantial decline in performance when faced with complex scenarios like stereo super-resolution. While DASR achieves the highest PSNR, it also demonstrates notable gaps in SSIM, LPIPS, and MADE compared to our proposed method, especially in terms of MADE, which effectively reflects the disparity consistency in stereo super-resolution. Compared to DASR with a MADE value of 3.01, our proposed method achieves a MADE value of 2.20, representing a difference of nearly 0.8.

Additionally, \cref{fig:flickr1024RS} illustrates the qualitative evaluation results of the above methods. It is clearly observed that, compared to other methods, the proposed method generates enhanced images with more realistic and rich textures, particularly in the case of images containing plants where our method performs more effectively. In order to visually compare the disparity differences before and after super-resolution, we utilized RAFR \cite{lipson2021raft} to estimate the disparity for all the super-resolved images, yielding disparity maps for each method. As shown in \cref{fig:flickr1024RSDis}, significant errors can be observed in the disparity maps obtained by real-world single-image super-resolution methods. However, the proposed method effectively preserves disparity consistency, demonstrating that our method not only enhances visual quality effectively but also maintains disparity consistency.

\textbf{StereoWeb20 Dataset.} To further validate the performance of our proposed method on a real dataset, we also evaluated it on the StereoWeb20 dataset. Since ground truth (GT) was not available, we relied on non-reference image quality evaluation metrics, such as NRQM and PI. The final evaluation results can be found in Table \ref{tab:stereoweb20}. Based on the results in the table, our method achieved the highest NRQM score and the second-best PI score.

Furthermore, \cref{fig:StereoWeb20} presents the qualitative evaluation results of the different methods on the StereoWeb20 dataset. It is evident that despite RealSR obtaining the highest PI score, the quality of the enhanced images is poor, featuring noticeable unnatural textures and some color shifts. In contrast, our proposed method, while not surpassing RealSR in terms of the PI score, produces enhanced images that are not only clear but also possess more realistic and natural textures, surpassing RealSR in terms of visual quality. Hence, these experiments effectively demonstrate the effectiveness of our proposed method.


\subsection{Ablation Study}
\textbf{Effectiveness of the Hybrid Degradation Strategy and Implicit Discriminator.} To investigate the effectiveness of the hybrid degradation strategy and implicit discriminator, an ablation study was performed using the NAFSSR\_L and SCGLANet stereo image generators on the Flickr1024RS dataset. The final results can be found in \cref{tab:hi}. From the table, it is evident that both the NAFSSR\_L and SCGLANet models, trained with the hybrid degradation strategy, effectively improve the PSNR and SSIM values, while significantly reducing MADE. Furthermore, the models trained with the implicit discriminator not only reduce LPIPS to enhance visual perception, but also further decrease the MADE values, thereby reducing the disparity error between the enhanced and original images.


\begin{table}[h]
\centering
\caption{Ablation results achieved on Flickr1024RS with/without hybrid degradation model and implicit discriminator. }
\label{tab:hi}
\resizebox{0.5\textwidth}{!}{
\begin{tabular}{lllllll}
\hline
Model &\begin{tabular}[c]{@{}l@{}}Hybrid \\ Degradation \end{tabular}  & \begin{tabular}[c]{@{}l@{}}Implicit\\ Discriminator\end{tabular}& PSNR $\uparrow$ & SSIM $\uparrow$ & LPIPS $\downarrow$ & MADE $\downarrow$ \\ \hline
NAFSSR\_L & $\times$  & $\times$  & 20.83 & 0.5686 & 0.4602 & 4.75 \\
 & \checkmark  & $\times$  & 21.84 & 0.6324 & 0.4159 & 2.77 \\
 & \checkmark  & \checkmark  & 20.88 & 0.5747 & 0.2231 & 2.54 \\ \hline
SCGLANet & $\times$  & $\times$  & 20.90 & 0.5739 & 0.4625 & 4.70 \\
 & \checkmark  & $\times$  & 22.01 & 0.6416 & 0.4082 & 2.48 \\
 & \checkmark  & \checkmark  & 21.06 & 0.6235 & 0.2098 & 2.20 \\ \hline
\end{tabular}
}
\end{table}
\begin{table}[h]
\centering
\caption{Ablation results achieved on Flickr1024RS trained with different data degradation strategies. }
\label{tab:dds}
\resizebox{0.45\textwidth}{!}{
\begin{tabular}{ccccccc}
\toprule
SO$^{*}$  & VB$^{*}$ & VN$^{*}$ & PSNR $\uparrow$ & SSIM $\uparrow$ & LPIPS $\downarrow$ & MADE $\downarrow$ \\
\midrule
$\times$     &    $\times$     &      $\times$         & 19.47 &0.5755  &0.2698  & 3.01             \\
\midrule
\checkmark     &    $\times$     &      $\times$          & 20.01 & 0.5867 & 0.2522 & 2.94             \\
$\times$     &    \checkmark     &      $\times$          & 19.61 & 0.5799 & 0.2601 & 2.84             \\
$\times$     &    $\times$     &      \checkmark          & 19.54 & 0.5781 & 0.2656 & 2.87              \\
\midrule
\checkmark     &    \checkmark     &      $\times$          & 19.98 & 0.5903 & 0.2524 & 2.73             \\
$\times$     &    \checkmark     &      \checkmark          & 19.89 & 0.5867 & 0.2507 & 2.68             \\
\checkmark     &    $\times$     &      \checkmark          & 20.03 & 0.5891 & 0.2486 & 2.70            \\
\checkmark     &    \checkmark     &      \checkmark          & 20.06 & 0.5893 & 0.2477 & 2.71            \\
\bottomrule
\multicolumn{7}{l}{$*$ SO, VB and VN represents shuffle operator, left-right vary blur kernel} \\
\multicolumn{7}{l}{\hspace{1.8mm} and vary noise level, respectively.}
\end{tabular}}
\vspace{-2mm}
\end{table}
\textbf{Impact of Different Data Degradation Strategies.} To investigate the impact of different data degradation strategies on model performance, this study performed a detailed exploration of various degradation strategies within the hybrid degradation model. This exploration included the second-order random degradation strategy (SO), various noise strategy (VN) for the left and right views, and various blur kernel strategy (VB) for the left and right perspectives. The final test results are presented in \cref{tab:dds}. It is evident from the table that individually incorporating the VB and VN strategies significantly reduces MADE, with the VB strategy showing particularly improved results. Moreover, while the inclusion of the SO strategy alone has a limited impact on reducing the MADE metric, it effectively improves image quality. Specifically, there is an improvement of 0.54 in PSNR and 0.0112 in terms of SSIM. Furthermore, by combining the SO strategy with either the VB or VN strategy, there is a further improvement in image quality and a significant reduction in disparity error.

\begin{figure}[!t]
\begin{center}
\scalebox{0.99}{
\begin{tabular}[b]{c@{ } c@{ } c@{ }  }

    \includegraphics[width=.20\textwidth,height=0.1\textheight,valign=t]{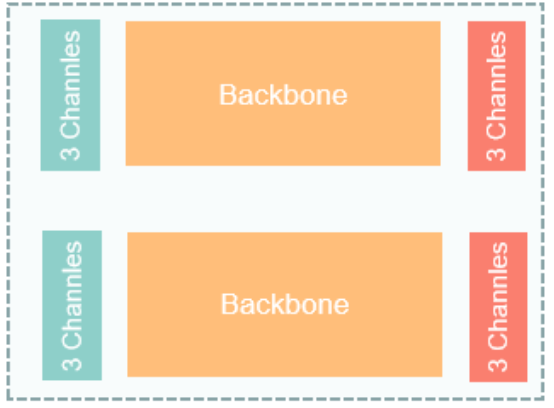} &
    \includegraphics[width=.20\textwidth,height=0.1\textheight,valign=t]{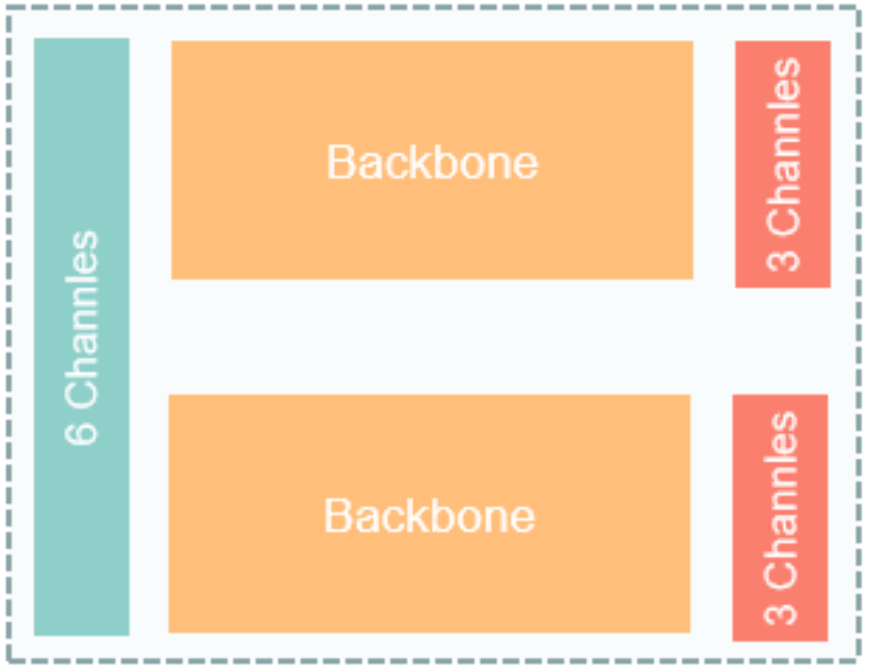} &  \\
    \scriptsize~(a) & \scriptsize~(b) &
    \\
    \includegraphics[width=.20\textwidth,height=0.1\textheight,valign=t]{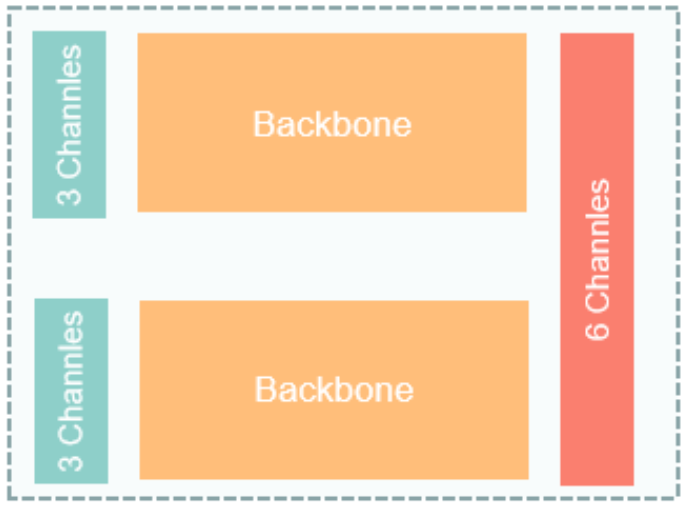} &
    \includegraphics[width=.20\textwidth,height=0.1\textheight,valign=t]{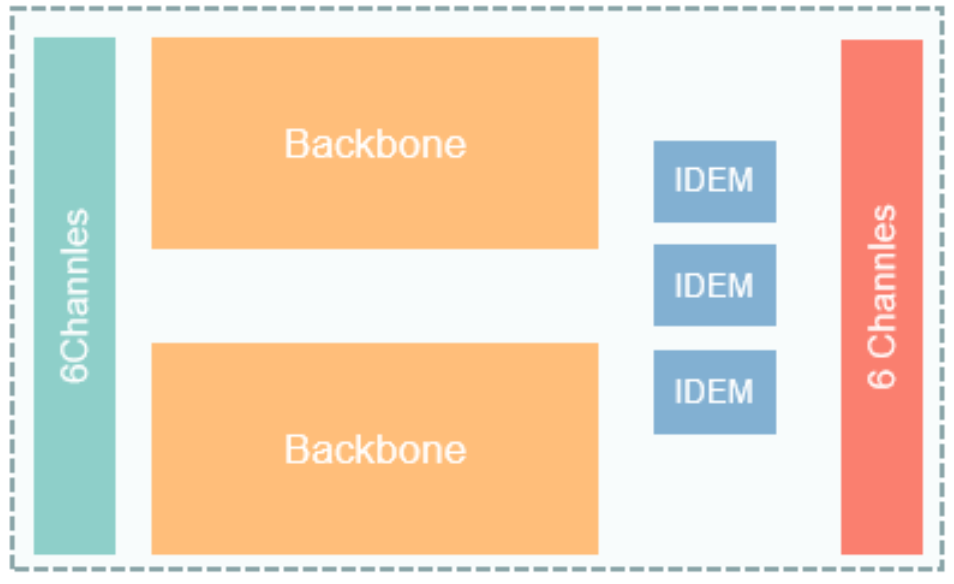} &  \\
    \scriptsize~(c) & \scriptsize~(d)

\end{tabular}}
\end{center}
\vspace{-1mm}
\caption{The architectures of four different discriminator.}
\label{fig:dis}
\vspace{-3mm}
\end{figure}

\textbf{Impact of Different Discriminators.} To investigate the effects of various discriminator types on model performance. The SCGLANet was used as the generator, and the results of models trained with different discriminator structures were compared. Specifically, four discriminator architectures are compared: 1) the baseline model with a single-image super-resolution discriminator, 2) a modification of the first convolutional layer of the single-image super-resolution discriminator from 3 channels to 6 channels, 3) the addition of a convolutional layer at the end of the single-image super-resolution discriminator to fuse the features of the left and right views, and 4) the incorporation of several IDEMs with 6-channel input and output. The differences among these four discriminator architectures are visually demonstrated in \cref{fig:dis}. The final evaluation results are presented in \cref{tab:dc}. The table reveals that changing the input channels of the discriminator from 3 to 6 and adding a convolutional fusion layer at the end effectively reduce the disparity error. This demonstrates that fusing the left and right views for joint judgment is the correct optimization direction for stereo image super-resolution. Furthermore, incorporating the multi-scale implicit information fusion module IDEM enables effective improvement in PSNR and SSIM while significantly reducing MADE. 

\subsection{Discussions and Limitations}
 Compared to single-image super-resolution in the real world, stereo-image super-resolution not only aims to enhance subjective visual effects but also seeks to minimize disparity errors between reconstructed pre- and post-images. This is the key distinction between stereo-image super-resolution and single-image super-resolution in the real world. Consequently, models designed for single-image super-resolution may not be entirely suitable for stereo-image super-resolution. Moreover, the detailed experiments in this paper confirm that models specifically tailored for stereo-image super-resolution in the real world are more effective for real-world stereo images than models developed for single-image super-resolution in the real world.

 \begin{table}
\centering
\caption{Ablation results achieved on Flickr1024RS trained with different discriminators.}
\label{tab:dc}
\begin{tabular}{ccccc}
\toprule
Methods$^{*}$  & PSNR $\uparrow$ & SSIM $\uparrow$ & LPIPS $\downarrow$ & MADE $\downarrow$ \\
\midrule
1          & 20.34 &0.6025  &0.2189  & 3.21             \\
\midrule
2         & 20.22 &0.5991  &0.2255  & 2.87              \\
3        & 20.29 &0.5966  &0.2140  & 2.64              \\
4        & 20.43 &0.6047  &0.2131  & 2.31               \\
\bottomrule
\multicolumn{5}{l}{ $*$ 1. Using the single-image discriminator (Baseline);} \\
\multicolumn{5}{l}{\hspace{1.6mm} 2. Converting the 3-channel input of Baseline to 6 channels;} \\
\multicolumn{5}{l}{\hspace{1.6mm} 3. Adding a convolutional fusion layer at the end of Baseline;} \\
\multicolumn{5}{l}{\hspace{1.6mm} 4. Using the proposed implicit disparity extraction module (IDEM)} \\
\multicolumn{5}{l}{\hspace{3.2mm}  and the convolutional fusion layer.}
\end{tabular}
\vspace{-2mm}
\end{table}

 Despite all this, there are specific challenges present in the field of stereo-image super-resolution when compared to single-image super-resolution in the real world. Firstly, there is a scarcity of high-quality stereo images. In the domain of super-resolution, having access to high-quality data is crucial for training high-quality super-resolution models. Additionally, due to the greater diversity and complexity of degradation in real-world stereo images when compared to single images, it becomes challenging to effectively handle all images solely through synthetic degradation schemes. When confronted with images that go beyond the degradation space, not only will the enhanced image quality suffer, but it will also result in disparity errors.

 Considering that single-image super-resolution has already matured as a field with established solutions and high-quality datasets, this paper suggests that pre-training models for single-image super-resolution can be efficiently transferred and fine-tuned for stereo-image super-resolution. One possible approach is to incorporate an adapter layer for fine-tuning. Additionally, exploring degradation models through contrastive learning to identify more universally applicable scenarios is also a promising direction for future research.

\section{Conclusion}

This paper presents a method for stereo image super-resolution in real-world scenarios. The proposed approach combines a realistic data hybrid degradation model and a stereo implicit discriminator to effectively enhance stereo images while preserving their inherent disparity. The design includes a three-stage data degradation model capable of handling the complexity and diversity of stereo images. Additionally, by embedding disparity information implicitly into the discriminator, the generator is compelled to generate high-quality stereo images without introducing disparity shifts. The effectiveness of the proposed method has been evaluated through comprehensive comparisons on both the synthetic and the real datasets. Results show that the proposed method achieves superior visual effects on the Flickr1024RS synthetic dataset while maintaining a low disparity error of only 2.2 between pre-enhanced and post-enhanced images, surpassing other real-world single-image super-resolution methods. Furthermore, competitive results are obtained on both qualitative and quantitative evaluations using the StereoWeb20 real dataset. Future research will focus on efficiently transferring knowledge from the single-image super-resolution domain to enhance its applicability to stereo super-resolution. Additionally, there will be continued exploration of more general stereo image degradation models. 


\bibliographystyle{ieeetr}
\bibliography{BibTexInfo}
\end{document}